\documentclass[aps,prl,preprint,groupedaddress]{revtex4-2}
\usepackage{graphicx}
\usepackage{xcolor}
\usepackage{epstopdf}
\usepackage{color,soul}
\usepackage[utf8]{inputenc}
\usepackage{hyperref}
\usepackage{lineno}
\usepackage{amsmath, bm}
\usepackage{mathtools}
\usepackage{upgreek}
\usepackage[normalem]{ulem}
\usepackage{lineno}
\begin{document}
	
\title{Tuning of anomalous magnetotransport properties in half-Heusler topological semimetal GdPtBi}
\author{Orest Pavlosiuk$^{1,*}$} 
\author{Piotr Wiśniewski$^{1}$} 
\author{Romain Grasset$^2$}
\author{Marcin Konczykowski$^2$}
\author{Andrzej Ptok$^3$} 
\author{Dariusz Kaczorowski$^{1,*}$} 
\affiliation{$^1$~Institute of Low Temperature and Structure Research, Polish Academy of Sciences, Ok\'{o}lna 2, 50-422 Wroc{\l}aw, Poland\\
$^2$~Laboratoire des Solides Irradiés, École Polytechnique, 91128 Palaiseau, France\\	 
$^3$~Institute of Nuclear Physics, Polish Academy of Sciences, W. E. Radzikowskiego 152, PL-31342 Krak\'{o}w, Poland\\
$^*$~Corresponding authors: d.kaczorowski@intibs.pl, o.pavlosiuk@intibs.pl}
	
\begin{abstract}
Half-Heusler compounds from the $RE$PtBi family exemplify Weyl semimetals in which external magnetic field induce Weyl nodes. 
These materials exceptionally host topologically non-trivial states near the Fermi level and their manifestation can be clearly seen in the magnetotransport properties. 
In this study, we tune the Fermi level of the archetypal half-Heusler Weyl semimetal GdPtBi through high-energy electron irradiation, moving it away from the Weyl nodes to investigate the resilience of the contribution of topologically non-trivial states to magnetotransport properties. 
Remarkably, we observe that the negative longitudinal magnetoresistance, which is a definitive indicator of the chiral magnetic anomaly occurring in topological semimetals, persists even when the Fermi level is shifted by 100\,meV from its original position in the pristine sample. 
Additionally, the anomalous Hall effect shows complex variations as the Fermi level is altered, attributed to the energy-dependent nature of the Berry curvature, which arises from avoided band crossing. 
Our findings show the robust influence of Weyl nodes on the magneto-transport properties of GdPtBi, irrespective of the Fermi level position, a behaviour likely applicable to many half-Heusler Weyl semimetals.
\end{abstract}
\maketitle
\section{Introduction}

The exceptional physical properties of topological materials makes them very promising from the perspective of their possible applications in quantum computing and spintronics.\cite{Hasan2010a,Armitage2018,Hasan2017} 
According to theoretical predictions, thousands of materials from the Inorganic Crystal Structure Database\cite{Hellenbrandt2004} have a non-trivial topology of electronic structure.\cite{Zhang2019g,Vergniory2019,Tang2019a} 
However, there is a much smaller number of experimentally confirmed topological materials.
Furthermore, there are only a few dozen materials demonstrating some features of these states in electronic transport properties. 
This huge difference in numbers seems to be caused by the fact that (i) topologically non-trivial states are located far below or above the Fermi level and do not contribute to electron transport; (ii) topologically non-trivial states coexist with trivial ones in the vicinity of the Fermi level and the latter dominate electron transport.  
Several half-Heusler phases with $RE$PtBi ($RE$-rare earth element) composition form a group of materials that overcome the above limitations and exhibit extraordinary transport properties that stem from topologically non-trivial states.\cite{Mun2022} 
One such compound is GdPtBi, an archetypal topological Weyl semimetal in which Weyl nodes are induced by magnetic field.\cite{Hirschberger2016a,Shekhar2018} 
Half-Heulsler antiferromagnet GdPtBi (N\'eel temperature, $T_N\!\approx\!9$\,K)  has been reported to demonstrate negative longitudinal magnetoresistance (NLMR)\cite{Hirschberger2016a,Shekhar2018}, planar Hall effect (PHE)\cite{Kumar2018} and anomalous Hall effect (AHE)\cite{Suzuki2016,Shekhar2018}, all stemming from topologically non-trivial states. 
The mechanism for the formation of magnetic field induced Weyl nodes in half-Heusler compounds is still debated in the literature, its source could be Zeeman splitting\cite{Hirschberger2016a} or magnetic exchange interactions.\cite{Shekhar2018} 
Alongside GdPtBi, several other rare earth-based half-Heusler materials have been reported to demonstrate features associated with a chiral magnetic anomaly,\cite{Pavlosiuk2019,Pavlosiuk2020,Chen2020a,Guo2018,Chen2021h} which is the fingerprint of any topological semimetal.\cite{Son2013} 
As it has been reported, the magnitude of NLMR\cite{Hirschberger2016a,Chen2020a} and AHE\cite{Zhu2023b} varies in the $RE$PtBi series with the change of $RE$, and even changes strongly from sample to sample for the same material. 
However, there are no systematic studies on how the magnitude of NLMR and AHE depends on the Fermi level position with respect to topologically non-trivial states. 
In their recent work, Sun et al.\cite{Sun2021} explored how AHE, MR and electronic structure depend on external pressure. 
They found that the AHE disappears at sufficiently high pressure, while the NLMR is almost unaffected by external pressure. 
Apart from the application of external pressure, the Fermi level can be tuned with other techniques like chemical doping or electrical gating. 
On the other side, in our recent work, we have successfully used electron irradiation to tune the Fermi level of another half-Heusler material LuPdBi, which is a mixed-parity superconductor.\cite{Ishihara2021a}  
Moreover, this technique has also been utilized in the study of magneto-transport in magnetic topological insulator.\cite{Sitnicka2023} 

Here, we describe a thorough study of the magneto-transport properties (Hall effect, transverse and longitudinal MR) of pristine and electron-irradiated samples of half-Heusler Weyl semimetal GdPtBi. 
The experimental results are accompanied by the results of theoretical calculations of electronic structure and anomalous Hall effect. 
Our findings indicate that with the Fermi level tuned away from the Weyl nodes, NLMR becomes smaller, and the magnitude of AHE and the magnetic field at which the AHE occurs change in a complex way, which is in accordance with the results of our theoretical calculations. 

\section{Results and discussion}

\subsection{Electrical resistivity}

\begin{figure}[h]
	\includegraphics[width=0.49\textwidth]{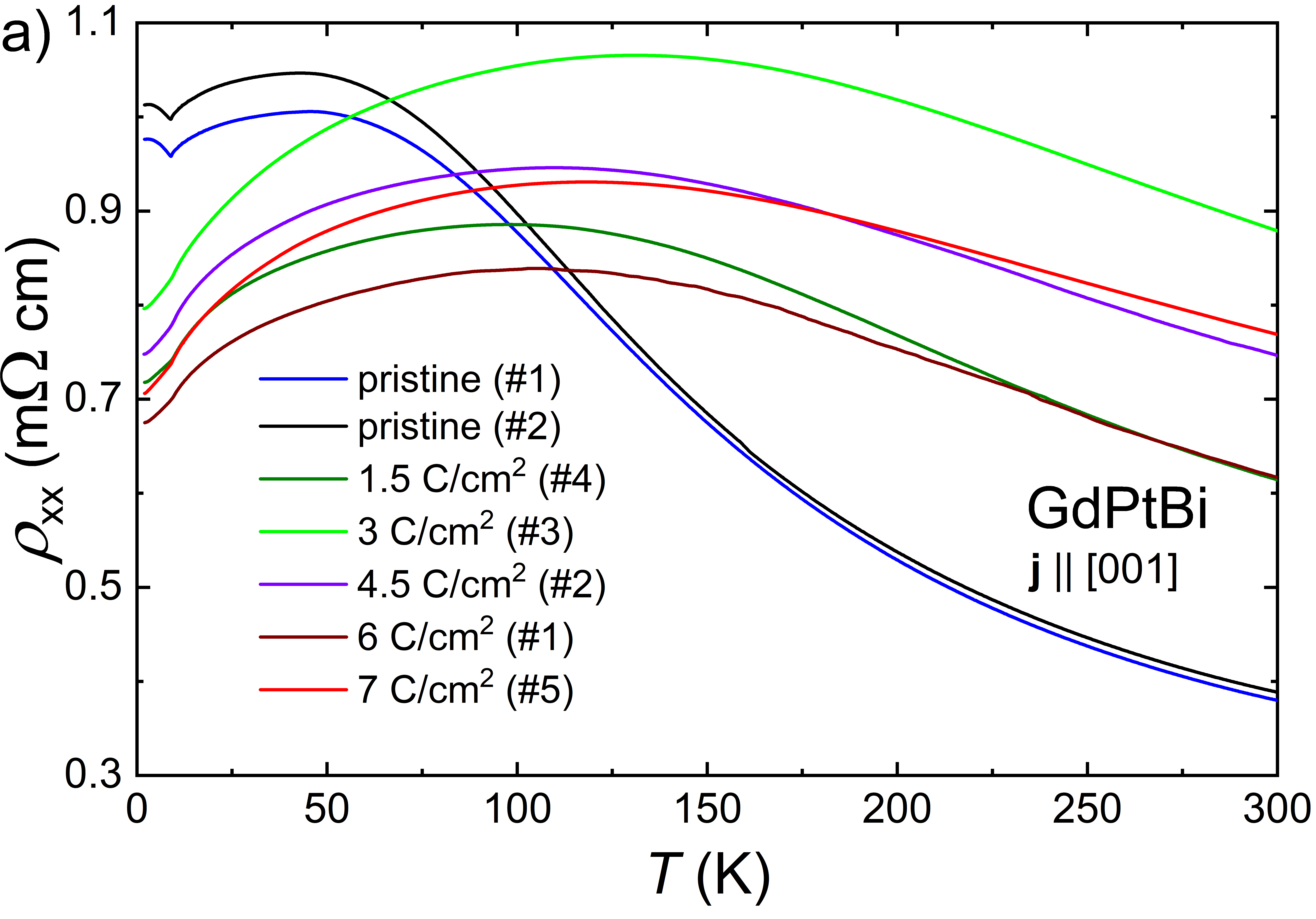}
	\includegraphics[width=0.49\textwidth]{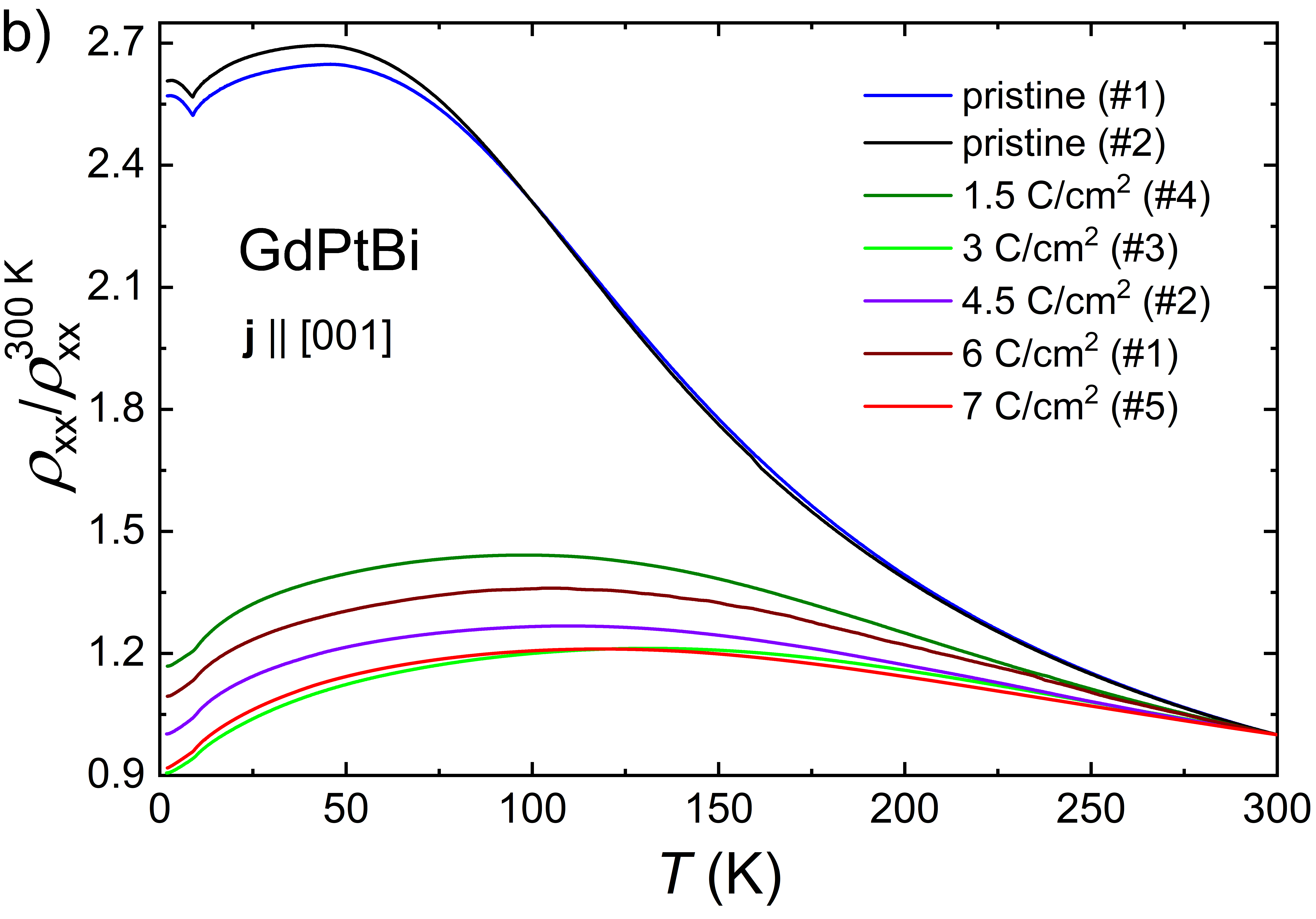}
	\caption{Temperature dependence of the electrical resistivity (a) and normalised electrical resistivity to the value of $\rho$ recorded at $T=300$\,K (b) of pristine and irradiated GdPtBi samples, measured with an electrical current (\textbf{j}) applied along [001] crystallographic direction and in zero magnetic field. 
		\label{rho}}
\end{figure}

The temperature ($T$) dependence of the electrical resistivity ($\rho$) of both pristine samples is almost identical (blue and black curves in Fig.~\ref{rho}a) and resembles the results reported in literature.\cite{Hirschberger2016a,Suzuki2016,Shekhar2018} 
This fact, in conjunction with the observed similarities in other magnetotransport properties (see below) for the pristine samples $\#1$ and $\#2$, allows us to assume that all pristine samples, $\#1\!-\!\#5$, have the same electron transport properties as they were all cut from the same single crystal. 
At higher temperatures, $\rho(T)$ demonstrates a semiconducting-like behaviour, which changes to metallic-like at lower $T$. 
There is also a pronounced valley in $\rho(T)$ of pristine samples in the vicinity of $T_N$, and an increase in $\rho$ at $T<T_N$ is associated with the formation of magnetic superzone gap.\cite{Mun2016a}
Interestingly, the electron-irradiated samples do not exhibit this valley in $\rho(T)$ and their $\rho$ does not increase as $T$ decreases; instead, $\rho$ continues decreasing. 
A similar impact on the shape of $\rho(T)$ close to $T_N$ has the external pressure, as reported in Ref.\cite{Sun2021} and has been attributed to the increase in hole concentration.   
In addition, at higher $T$, electron irradiation also leads to changes in $\rho(T)$, but these do not show a monotonic behaviour with a gradual increase in the irradiation dose, $\phi$. 
For all irradiated samples, the maxima in $\rho(T)$ (associated with the border between semiconducting- and metallic-like behaviour) shift to higher $T$ (samples become more metallic). 
In order to eliminate measurement error in the sample size that might affect $\rho$ values, we normalised $\rho(T)$ dependencies to the $\rho$ value at $T=300$\,K ($\rho(T)/\rho(300\,K)$), as shown in Fig.~\ref{rho}b. 
The effect of electron irradiation on $\rho(T)$ of GdPtBi differs significantly from that described for metallic materials, in which $\rho(T)$ shifts upward with increasing irradiation dose.\cite{Mizukami2014} 
The observed behaviour of $\rho(T)$ for GdPtBi samples also confirmed the fact that irradiation not only provides the additional carriers, but also enhances the impurity scattering, in line with the previous report for LuPdBi.\cite{Ishihara2021a}         

\subsection{Hall effect}  

\begin{figure*}
	\includegraphics[width=0.49\textwidth]{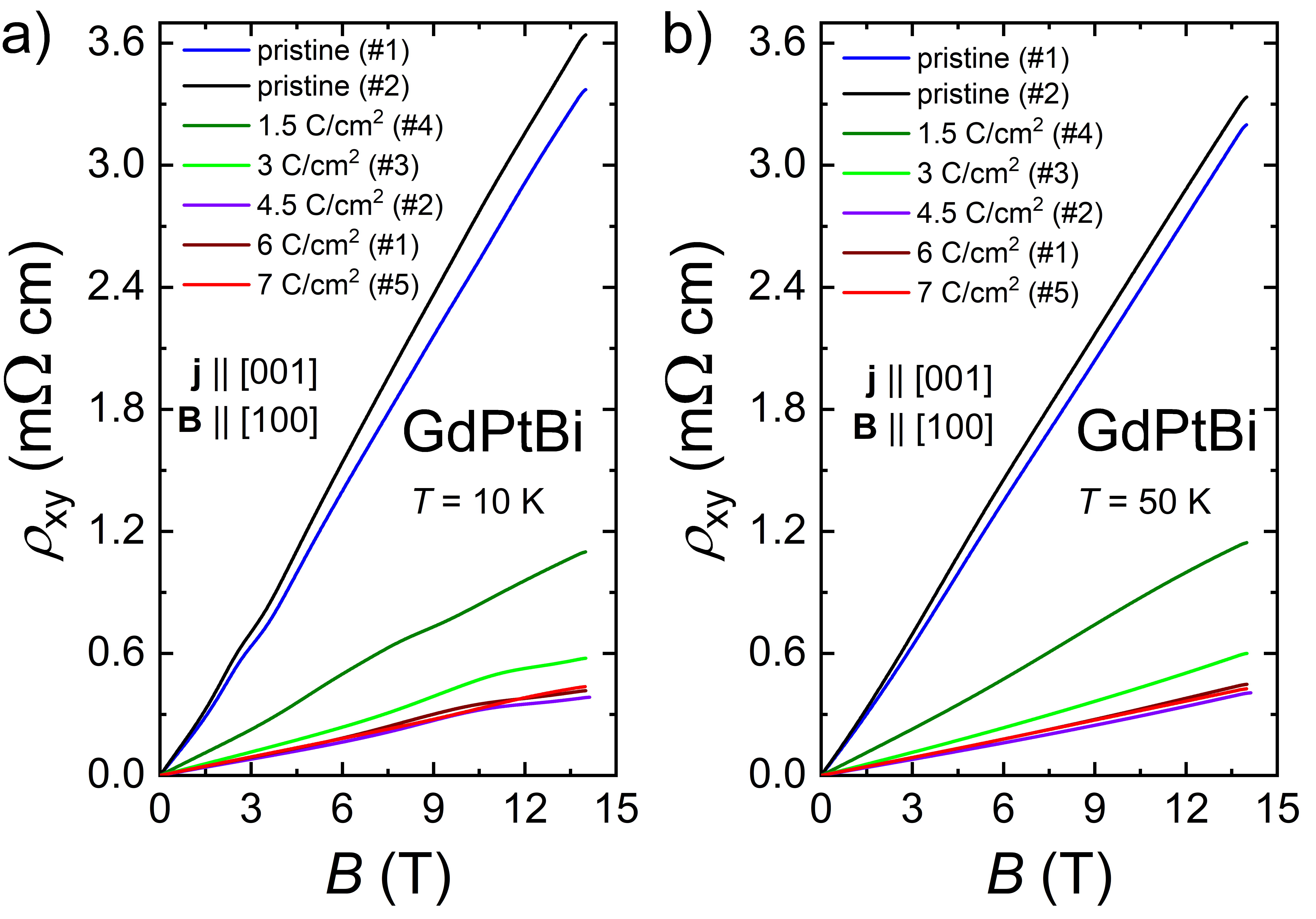}
	\includegraphics[width=0.49\textwidth]{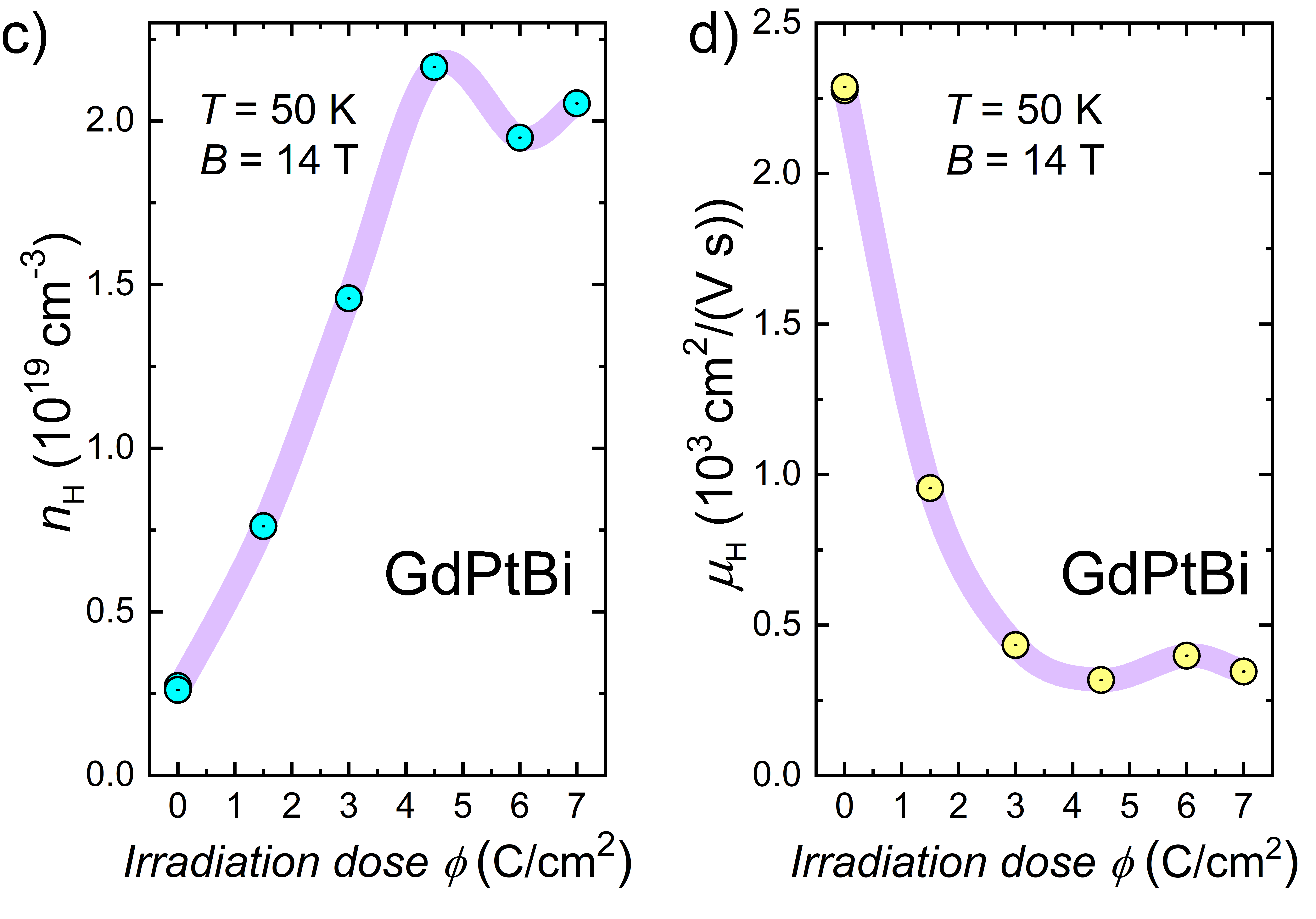}
	\caption{Magnetic field dependence of Hall electrical resistivity at $T=10$\,K (a) and $T=50$\,K (b). Carrier concentration (c) and carrier mobility (d) as a function of irradiation dose at $T=50$\,K.
	\label{Hall}}
\end{figure*}

To quantify the change in carrier concentration and mobility induced by electron irradiation, we have studied the Hall effect (see Fig.~\ref{Hall}). 
Figs.~\ref{Hall}a and \ref{Hall}b show the magnetic field ($B$) dependence of Hall resistivity ($\rho_{xy}$) for all samples studied at $T=10$\,K and 50\,K, respectively. 
At both values of $T$, the magnitudes of $\rho_{xy}$ are almost the same for a given sample, meaning that the concentration and mobility of carriers barely change with a variation of $T$. 
The most striking difference is that at $T=10$\,K the contribution of anomalous Hall effect (AHE) is pronounced, while at $T=50$\,K it is absent (see discussion below). 
In order to omit uncertainty in the determination of $n_H$ and $\mu_H$ caused by the presence of AHE, we calculated both parameters using the data recorded at $T=50$\,K. 
It should be noted that for all studied samples, hole-type carriers dominate the electron transport, as previously reported in literature. 
Most of $RE$PtBi have the same type of conductivity.\cite{Hirschberger2016a,Suzuki2016,Shekhar2018,Pavlosiuk2019,Pavlosiuk2020,Pavlosiuk2016b,Zhang2020k} 
Both pristine samples demonstrate almost identical behaviour of $\rho_{xy}(B)$, indicating that their electron transport parameters are the same, the slight difference may be due to the different orientation of the samples with respect to the applied current and/or magnetic field, or to an error in the estimation of the sample thickness.
Electron irradiation also leads to the decrease in $\rho_{xy}$ values, similar to LuPdBi.\cite{Ishihara2021a}  
However, in contrast to LuPdBi, $\rho_{xy}$ of GdPtBi saturates at $\phi\geq4.5$\,C/cm$^2$. 
For all studied samples, $\rho_{xy}(B)$ dependences are almost rectilinear, except for the $B$ regions where AHE was observed (see below). 
This implies that the electron transport is dominated by carriers of one type. 
Therefore, we used the single-band approximation to determine the carrier concentration ($n_H$), the same model has been used in several works related to GdPtBi\cite{Hirschberger2016a,Shekhar2018,Sun2021} and other $RE$PtBi,\cite{Mun2016a} which allows us to better compare our data with literature data. 
We calculated $n_H$ using the formula: $n_H=B/(e\rho_{xy})$.
The obtained $n_H(\phi)$ dependence for the Hall effect data collected at $T=50$\,K and in $B=14$\,T is shown in Fig.~\ref{Hall}c, at lower $\phi$, $n_H$ increases with increasing irradiation dose up to 4.5\,C/cm$^2$, at higher doses $n_H$ is almost constant. 
This behaviour is different from that observed for LuPdBi\cite{Ishihara2021a}, where $n_H(\phi)$ does not saturate but increases over the entire range of studied $\phi$. 
However, a similar kind of saturation has been reported for electron-irradiated Mn-doped BiSeTe$_3$ in Ref.~\cite{Sitnicka2023}. 
The saturation can be understood as follows: the rate of electron removal from the band is equal to the defect creation rate multiplied by the number of free electrons captured by one defect. 
In our case, if the Fermi energy is higher than the electronic level formed on the defect, we continue to deplete the valence band and the hole concentration increases. 
At some point, the Fermi energy reaches the electronic state, and newly created defects will no longer capture additional electron. 
The variation of hole concentration with dose will saturate at the level of Fermi energy, coinciding with the electronic defect. 

In the case of GdPtBi, the irradiation leads to an increase of $n_H$ by one order of magnitude from the value of $2.6\times10^{18}$\,cm$^{-3}$ for pristine sample to $2.2\times10^{19}$\,cm$^{-3}$ for the sample irradiated with 4.5\,C/cm$^2$. 
This increase in $n_H$ is considerably greater than that caused by external pressure ($n_H$ has been found to increase from $1.05\times10^{18}$\,cm$^{-3}$ to $1.28\times10^{18}$\,cm$^{-3}$ when a pressure of 2.1\,GPa was applied).\cite{Sun2021}
$n_H$ of pristine GdPtBi samples is of the same order of magnitude as that reported in the literature.\cite{Hirschberger2016a,Suzuki2016,Sun2021} 
Another parameter of electronic transport that can be derived from the Hall effect data is the carrier mobility ($\mu_H$), which was calculated using the formula $\mu_H=\rho_{xy}/(\rho_{xx}B)$. The dependence of $\mu_H$ on the irradiation dose is shown in Fig.~\ref{Hall}d. 
In the pristine samples, the mobility is close to $\sim\!2600$\,cm$^2$/(Vs), and as the irradiation dose increases $\mu_H$ decreases.
At $\phi\geq3$\,C/cm$^2$, $\mu_H$ is hardly affected by higher doses and only undergoes a slight change, reaching a value of $\sim\!350$\,cm$^2$/(Vs) for $\phi\geq7$\,C/cm$^2$.  
This behaviour is reminiscent of that reported for LuPdBi in Ref.\cite{Ishihara2021a}, but the saturation of mobility occurs at lower irradiation doses. 
The decreasing behaviour of $\mu_H(\phi)$ and the increasing behaviour of $n_H(\phi)$ can be understood as a synergy of two effects: (i) a shift of the Fermi level (Fermi pockets become larger, like their effective masses) and (ii) an increase in impurity scattering. 

\subsection{Magnetoresistance}

\begin{figure*}
	\includegraphics[width=0.49\textwidth]{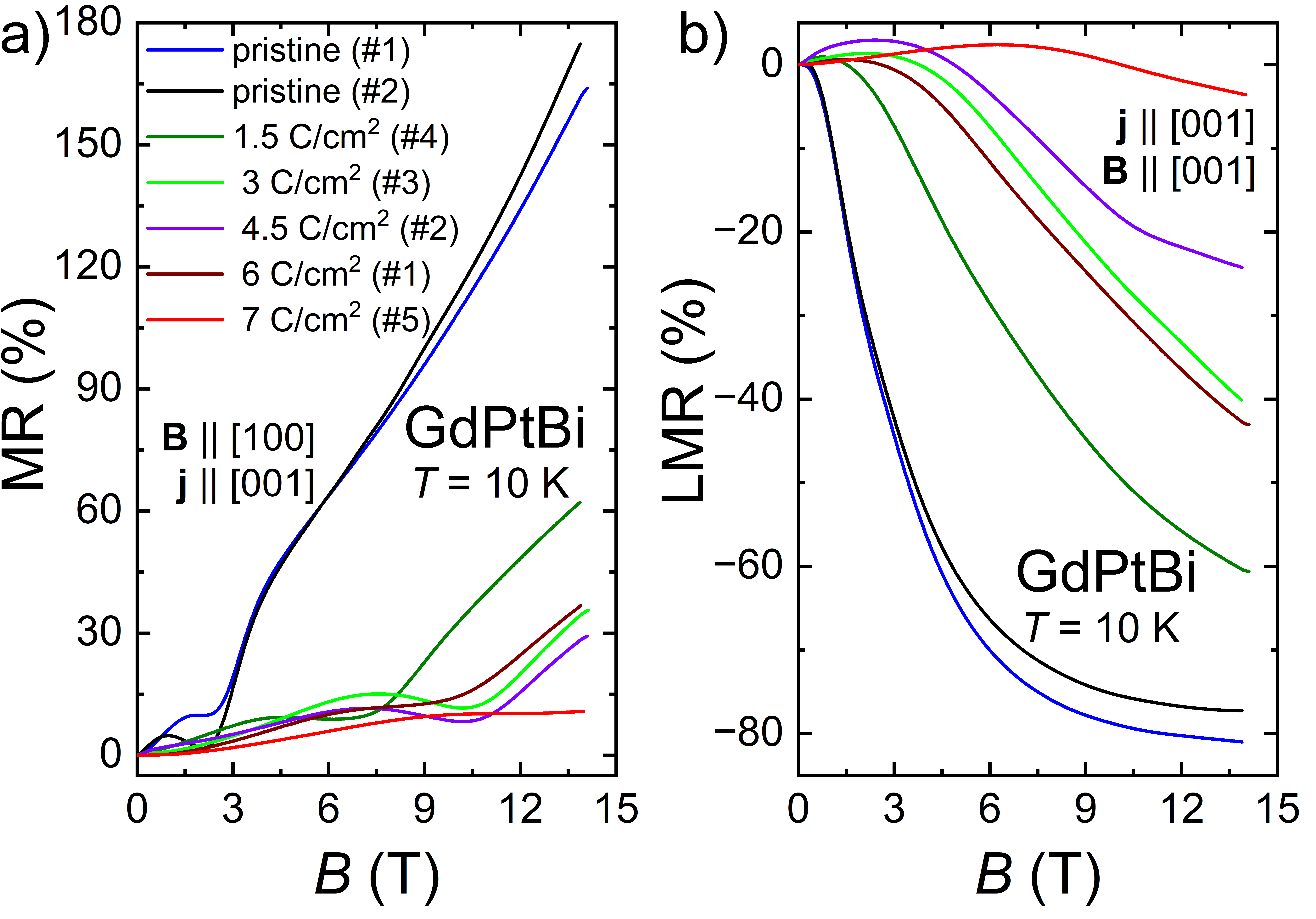}
	\includegraphics[width=0.49\textwidth]{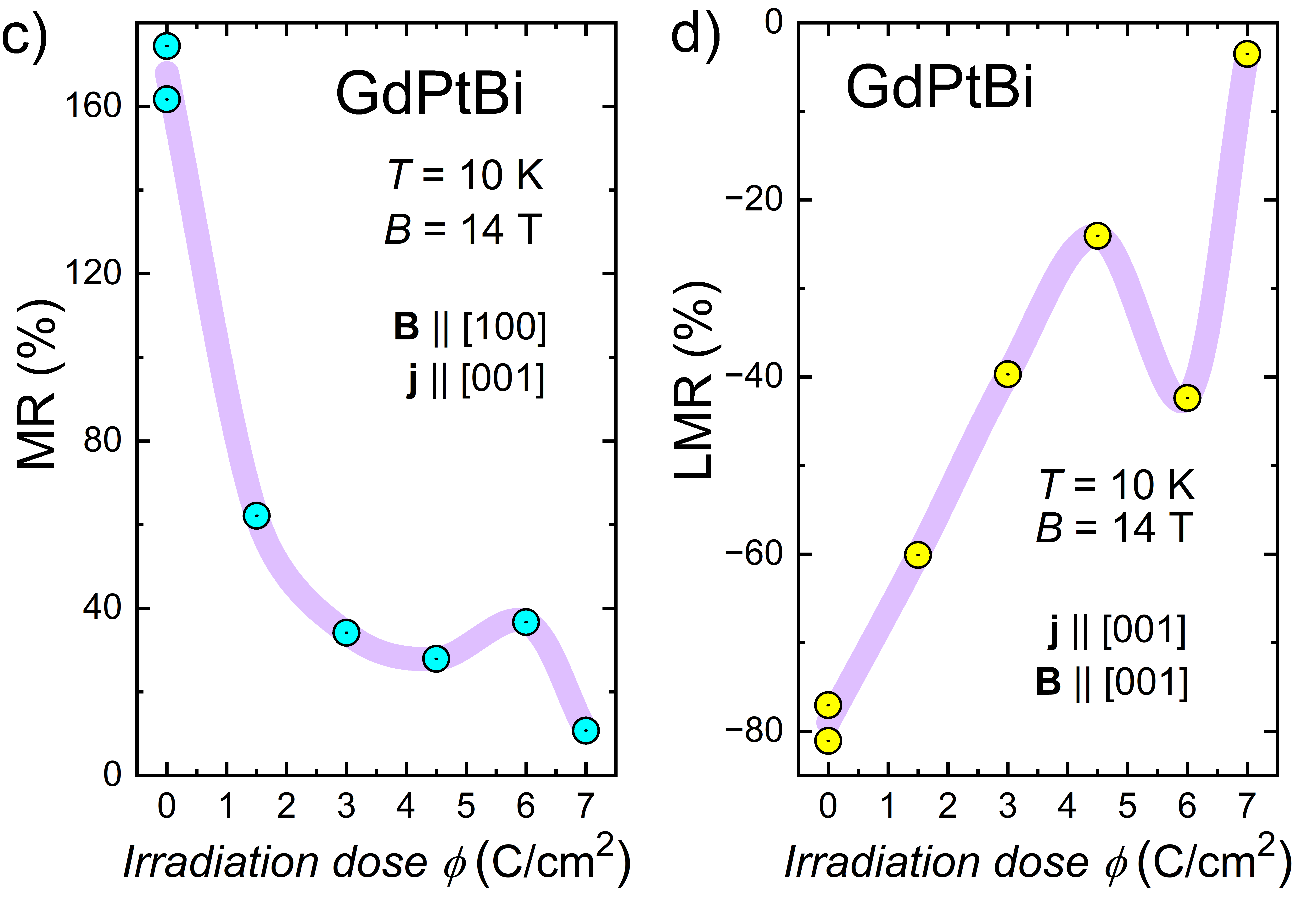}
	\includegraphics[width=0.49\textwidth]{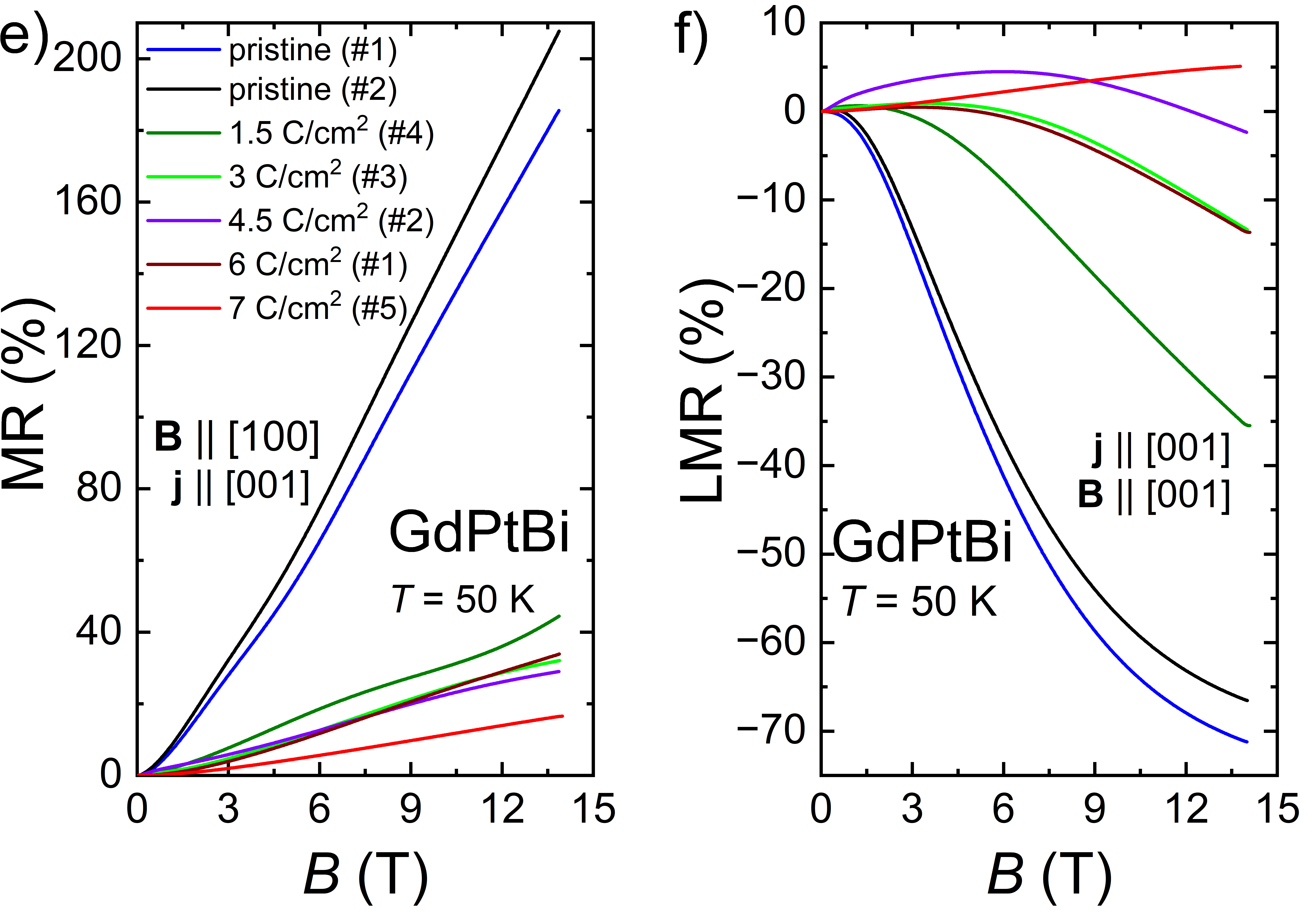}
	\includegraphics[width=0.49\textwidth]{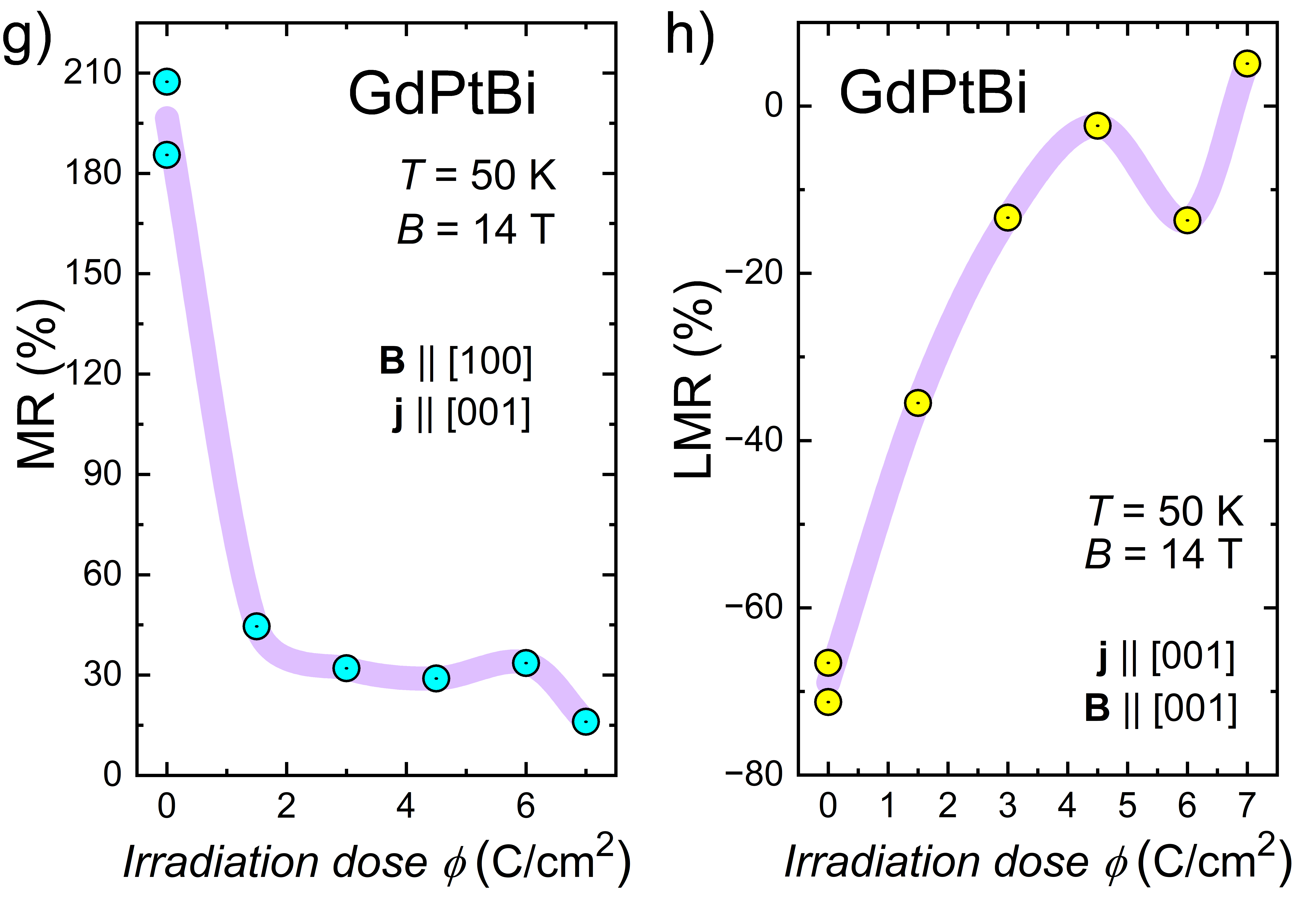}
	\caption{Transverse magnetoresistance of pristine and electron irradiated samples at $T=10$\,K (a) and $T=50$\,K (e). Longitudinal magnetoresistance of pristine and electron irradiated samples at $T=10$\,K (b) and $T=50$\,K (f). Transverse (c, g) and longitudinal (d, h) magnetoresistance as a function of irradiation dose recorded in $B = 14$\,T and at $T=10$\,K (c, d) and at $T=50$\,K (g, h). Violet lines in (c, d, g, h) are guides for the eye.
	\label{MR}}
\end{figure*}

The results of electrical resistivity and Hall effect measurements confirmed that electron irradiation induces a shift in the Fermi level in GdPtBi. 
Therefore, in the subsequent stage of our research, we investigated the impact of this shift on the magnetotransport properties, with a particular focus on its influence on the chiral magnetic anomaly, which has been previously reported for GdPtBi.\cite{Hirschberger2016a,Shekhar2018} 
The magnetoresistance (${\rm{MR}}=100\times[\rho_{xx}(B)/\rho_{xx}(B\!=\!0)-1]$) isotherms (measured in both the transverse ($\textbf{j}\perp\textbf{B}$) and longitudinal ($\textbf{j}\parallel\textbf{B}$) geometries) recorded at several different $T$ for the pristine samples $\#1$ and $\#2$ are shown in the Supplementary Materials (see Fig.~S1). 
The results obtained for both pristine samples are very similar to those reported in the literature.\cite{Hirschberger2016a,Shekhar2018,Sun2021} 
For both samples, the traverse MR does not saturate even in the strongest magnetic field. 
The nonmonotonous behaviour at low $B$ occurs in the same $B$ range, in which AHE was observed (see discussion below). 
In turn, for both pristine samples LMR is negative up to at least $T=100$\,K and becomes positive at higher $T$. 
Negative LMR can be the fingerprint of chiral magnetic anomaly (CMA) or may originate from a parasitic current jetting effect.\cite{Liang2018a}  
According to several studies, the latter effect can be excluded in GdPtBi.\cite{Hirschberger2016a,Liang2018a} 
We then studied the magnetotransport properties of irradiated samples, the results obtained at $T=10$\,K and 50\,K are summarized in Fig.~\ref{MR}(a-d) and Fig.~\ref{MR}(e-h), respectively.
Both temperatures are above the $T_N$, but at $T=10$\,K MR are affected by AHE, which is negligibly small at $T=50$\,K. 
At both $T$ values, magnitudes of TMR and negative LMR for the irradiated samples are smaller than for the pristine samples. 
The analysis of the behaviour of TMR($B$) at different $\phi$ at $T=10$\,K is complicated by fact that the maximum of AHE changes its magnitude and position with changing $\phi$ (see below). 
However, in a strong magnetic field at $T=10$\,K, it is evident that, as $\phi$ increases, the TMR decreases (see Fig.~\ref{MR}(a, c)), only with the exception of sample $\#1$ irradiated with 6\,C/cm$^2$, its TMR deviates from the overall trend. 
Interestingly, this particular sample exhibits a deviation from the general trends in all the transport properties studied here and in all extracted electron transport parameters. 
At $T=50$\,K and in the entire range of magnetic field, TMR gradually decreases with increasing $\phi$, with the exception of aforementioned sample $\#1$ irradiated with 6\,C/cm$^2$ (see Fig.~\ref{MR}e, g). 

\begin{figure}
	\includegraphics[width=0.24\textwidth]{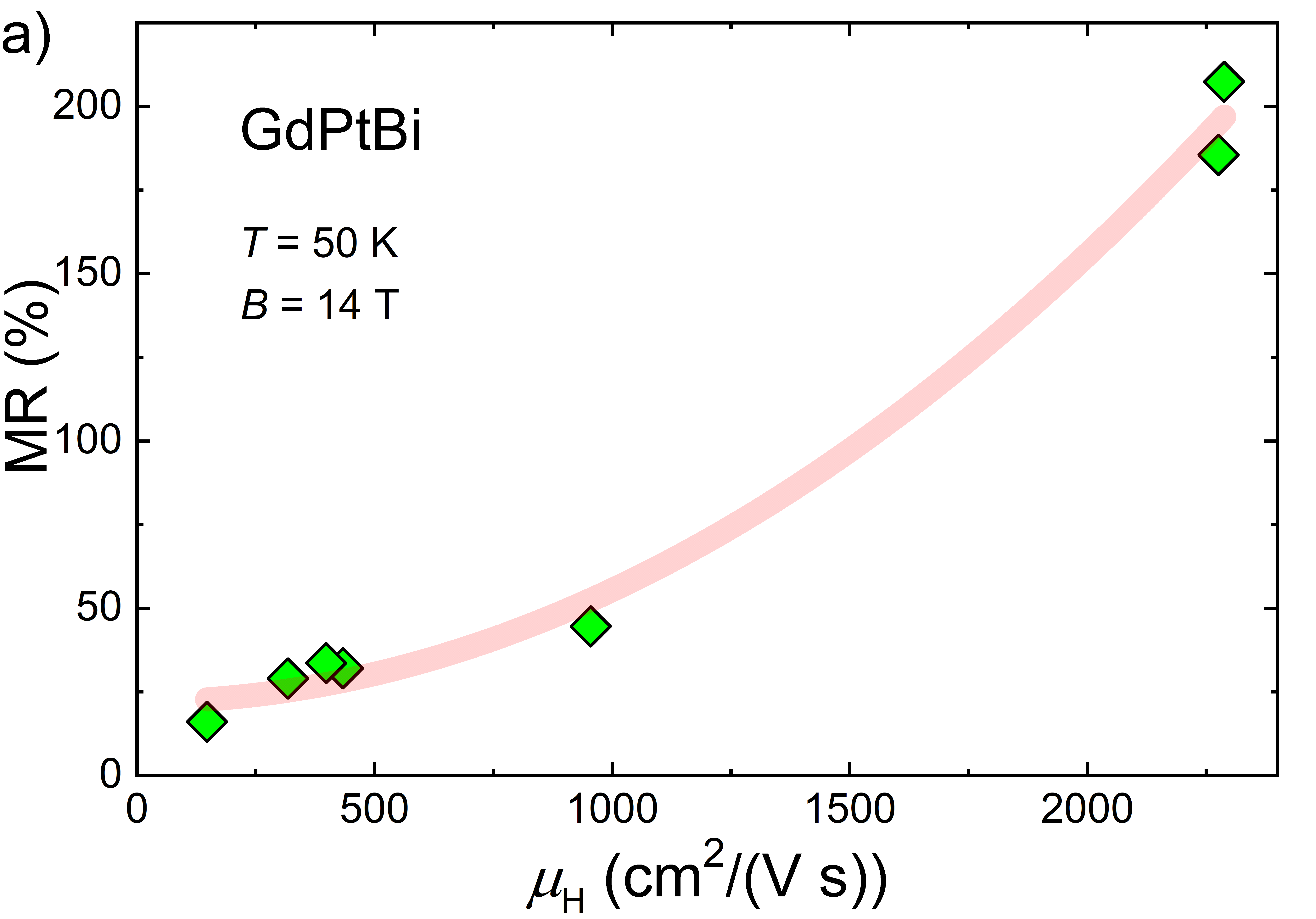}
	\includegraphics[width=0.24\textwidth]{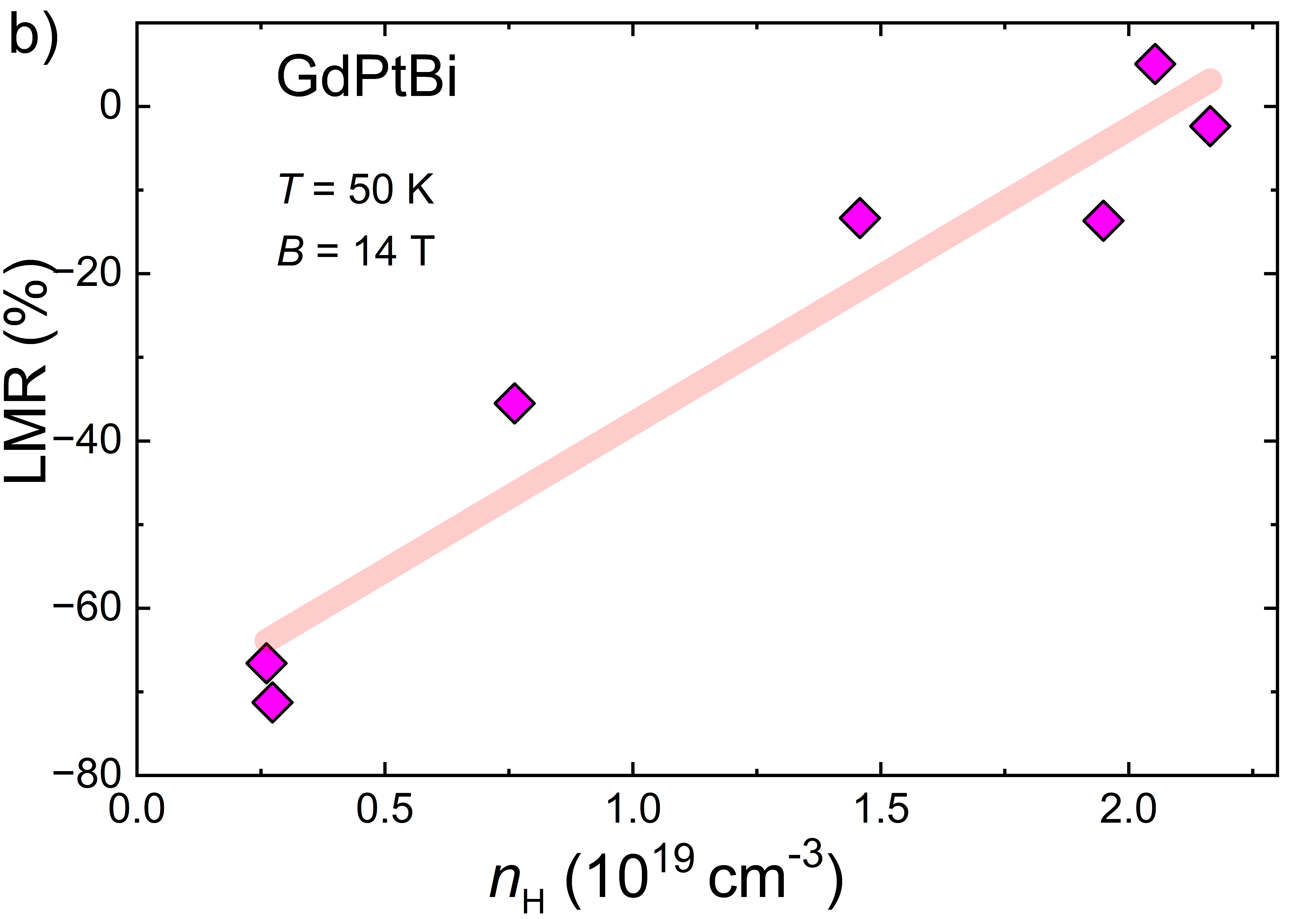}
	\caption{Transverse (a) and longitudinal (b) magnetoresistance of pristine and electron irradiated samples as a function of carrier mobility (a) and carrier concentration (b) at $T=50$\,K and in $B=14$\,T. Pink lines are guides for the eye.
		\label{Trends}}
\end{figure}

From the perspective of how the contribution of Weyl states to electron transport changes with Fermi level tuning, measurements of LMR are the most informative. 
We investigated LMR($B$) of all pristine and irradiated samples at $T=10$\,K and $T=50$\,K (see Fig.\,\ref{MR}b, f).
In half-Heuslers materials as well as in several other topological semimetals, the total LMR is a sum of two contributions, positive and negative.\cite{Huang2015,Pavlosiuk2020,Guo2018} 
The positive contribution may originate from a weak antilocalization effect\cite{Hikami1980} and/or Fermi surface anisotropy.\cite{Pal2010} The negative contribution is associated with the chiral magnetic anomaly.\cite{Son2013} 
Our findings indicate that the contribution of the latter to the overall LMR of GdPtBi is more pronounced at lower temperatures. 
However, as the irradiation dose increases, this contribution becomes smaller. 
For both pristine samples in $B=14$\,T, LMR$\sim\!-80\%$ and LMR$\sim-\!70\%$ at $T=10$\,K and 50\,K, respectively. 
Upon irradiation, LMR gradually decreases (with the only exception of sample $\#1$ irradiated with 6\,C/cm$^2$). 
At $T=10$\,K in the entire range of $\phi$, LMR is negative in the strongest magnetic fields, whereas in weak magnetic fields, LMR is positive.  
However, the region in which LMR is positive increases with $\phi$. 
At $T=50$\,K, the situation is quite similar with the only difference that LMR is smaller and for sample irradiated with 7\,C/cm$^2$, LMR is positive throughout the entire range of $B$ studied. 
It has been reported in many topological semimetals\cite{Pavlosiuk2020,Guo2018,Pavlosiuk2021,Huang2015,Lv2017} that with increasing $T$, CME becomes less pronounced due to the increasing phonon scattering with temperature.\cite{Zhang2016} 

A comparison of the magnetoresistance and Hall effect data revealed two interesting trends. 
First, transverse MR increases with increasing carrier mobility, and second, LMR increases with increasing carrier concentration. 
Transverse MR dependence is quadratic, which corresponds to the semi-classical model according to which MR$\sim\mu_H^2$ for low-carrier systems.\cite{Ali2015,Ishihara2021a} 
In turn, LMR is found to depend on $n_H$ almost linearly. 
Previously, a similar behaviour of LMR($n_H$) was reported in Ref.\,\cite{Hirschberger2016a}, however, in our investigations, we extended the range of $n_H$ considerably and did this in a more controllable way by gradually changing the irradiation dose.    

\subsection{Anomalous Hall effect} 

\begin{figure}
	\includegraphics[width=0.49\textwidth]{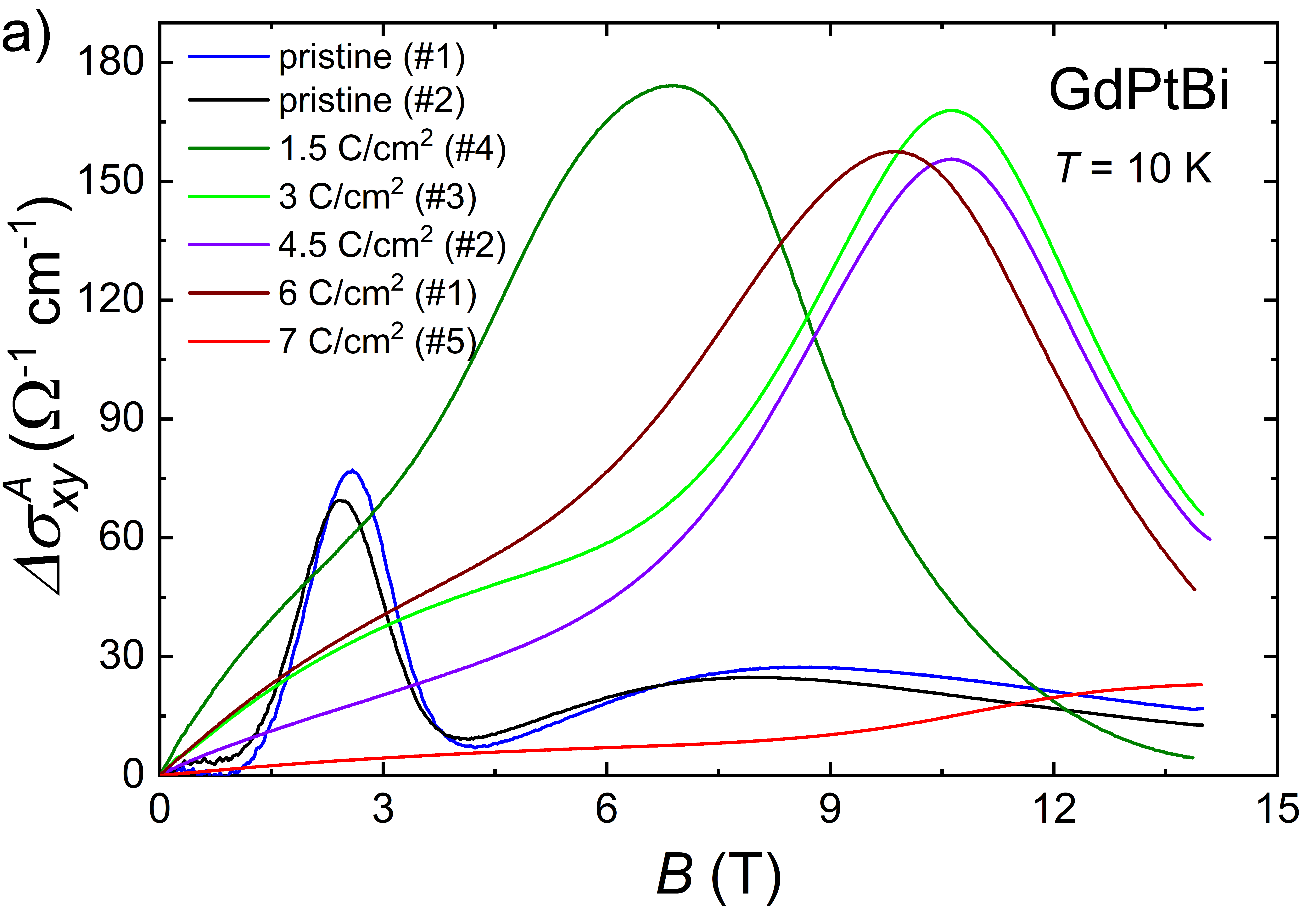}
	\includegraphics[width=0.49\textwidth]{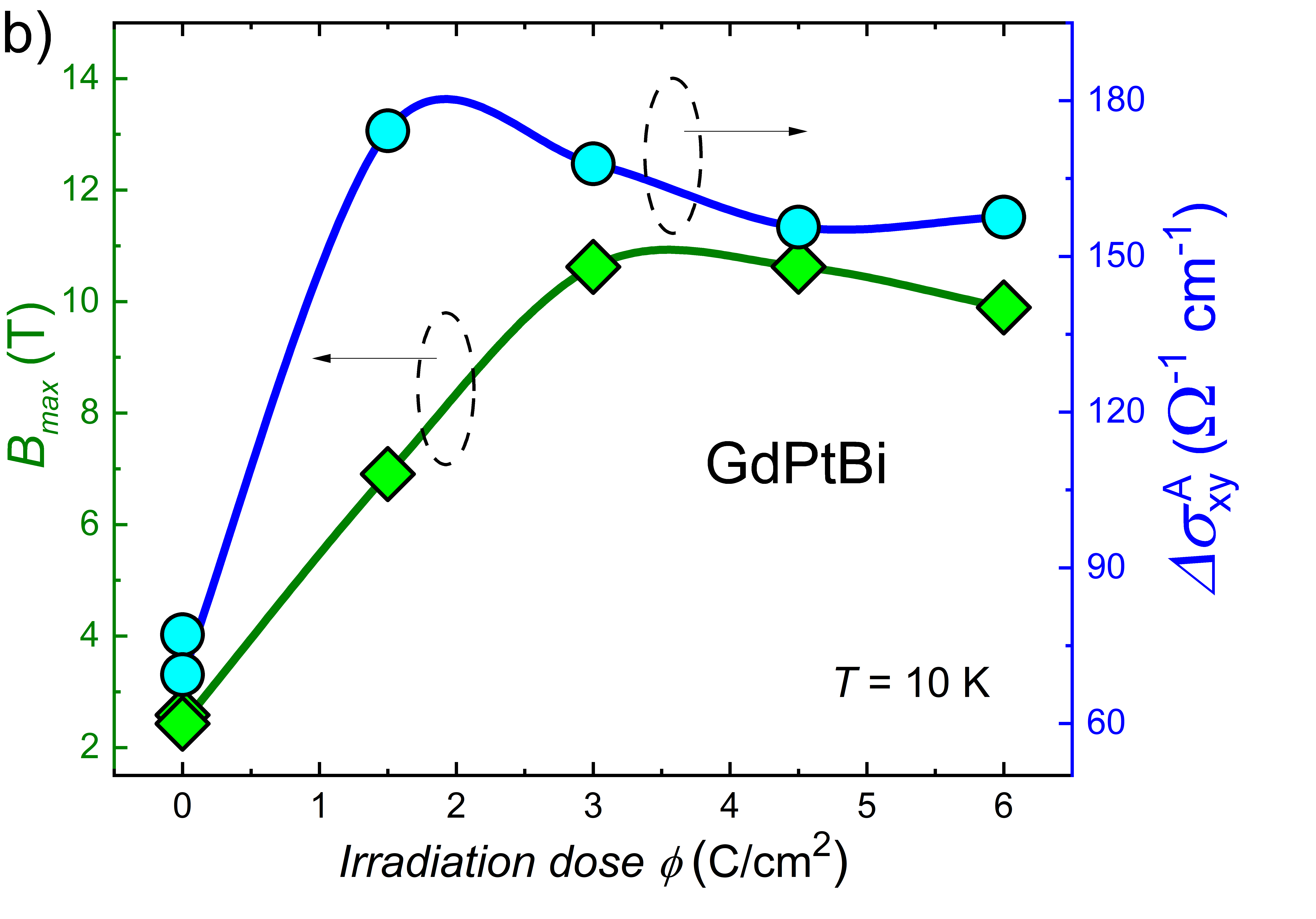}
	\caption{(a) Anomalous Hall conductivity as a function of magnetic field at $T=10$\,K for pristine and irradiated sample. (b) Maximum values of anomalous Hall conductivity (right axis) and magnetic field values (left axis) at which these maxima were recorded as a function of irradiation dose at $T=10$\,K.
		\label{AHE_summary}}
\end{figure}

\begin{figure*}
	\includegraphics[width=0.99\textwidth]{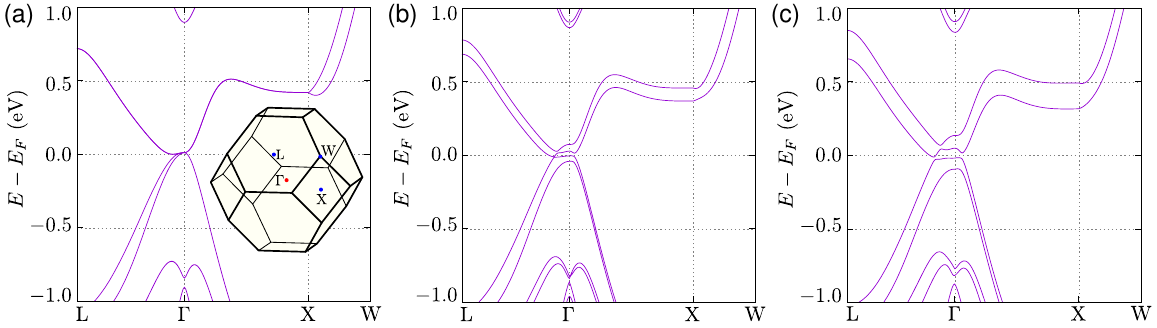}
	\caption{The electronic band structure in the absence (a) and presence of magnetic field of $5$~T (b) and $10$~T (c). The inset to panel (a) shows the bulk Brillouin zone with high symmetry points marked.
		\label{fig.band}}
\end{figure*}  

As mentioned in previous sections, anomalous Hall effect was observed in our pristine samples. 
Consequently, we also studied the impact of electron irradiation on AHE. 
GdPtBi was the first half-Heusler phase among the $RE$PtBi compounds where the observation of AHE has been reported.\cite{Suzuki2016} 
Following this, numerous studies on the observation of AHE in $RE$PtBi compounds were conducted.\cite{Pavlosiuk2020,Shekhar2018,Zhang2020k,zhu.singh.20} 
More recently, this study was extended to $RE$PdBi systems.\cite{Zhu2023b} 
The detailed procedure of the extraction of AHE contribution ($\Delta\sigma^A_{xy}$) is described in the Supplementary Materials and is the same as previously reported.\cite{zhu.singh.20} 
For both pristine samples, the difference in $\Delta\sigma^A_{xy}(B)$ curves is minimal across the entire range of magnetic fields covered. 
A minor discrepancy may be attributed to the missorientation of samples, as it has been shown recently that the AHE in HoPtBi is strongly anisotropic.\cite{Chen2021d} 
$\Delta\sigma^A_{xy}$ achieves the maximum values of $77.1\,\Omega^{-1}\rm{cm}^{-1}$ and $69.4\,\Omega^{-1}\rm{cm}^{-1}$ at $B_{max}=2.6$\,T and $B_{max}=2.4$\,T for pristine samples $\#1$ and $\#2$, respectively (see Fig.~\,\ref{AHE_summary}). 
These values are comparable to those reported for GdPtBi in Ref.\,\cite{Shekhar2018}.

It was observed that electron irradiation resulted in a pronounced change in the $\Delta\sigma^A_{xy}(B)$. 
Irradiation with the smallest dose of $\phi=1.5$\,C$/$cm$^2$ leads to a shift in the maximum of $\Delta\sigma^A_{xy}$ to the stronger magnetic field, $B_{max}=6.9$\,T, and also the magnitude of $\Delta\sigma^A_{xy}$ increases to the value of $174.2\,\Omega^{-1}\rm{cm}^{-1}$, the largest observed value of $\Delta\sigma^A_{xy}$ for any of the studied samples.  
As the irradiation dose is increased further, $B_{max}$ increases and saturates at value of $\sim10$\,T within the range of irradiation dose $3-6$\,C$/$cm$^2$. 
It appears that for the sample irradiated with $\phi=7$\,C$/$cm$^2$ the maximum in the $\Delta\sigma^A_{xy}(B)$ was not attained in the magnetic field up to 14\,T.    
The irradiation dose dependence of both $\Delta\sigma^A_{xy}$ and $B_{max}$ is shown in Fig.~\,\ref{AHE_summary}b. 

In order to gain insight into the complex behaviour of $\Delta\sigma^A_{xy}(\phi)$ and $B_{max}(\phi)$, we performed calculations of the electronic structure and anomalous Hall effect for several different values of magnetic field from the range 0-10\,T.        

\subsection{Electronic band structure calculations}

The electronic band structure of GdPtBi is presented in Fig.~\ref{fig.band} without and with the inclusion of the magnetic field.
In the absence of the magnetic field, the band structure is very similar to those previously reported for GdPtBi\cite{Suzuki2016, Shekhar2018} and other half-Heusler compounds.~\cite{souza.crivillero.23}
At $\Gamma$ point, band touching occurs, with the states around the Fermi level originating from Pt and Bi atoms (see Fig.~\ref{fig.band}a).
Additionally, inversion of $\Gamma_{8}$ ($p$-type) and $\Gamma_{6}$ ($s$-type) bands is realized.~\cite{Feng2010} 
The introduction of magnetic field leads to the lifting of band degeneracy and the emergence of avoided crossing features along high-symmetry directions.
Moreover, the appearing of the Fermi surface around $\Gamma$ point is observed.
The observed effect is similar to that reported for CePtBi,~\cite{kozlova.hagel.05} GdPtBi~\cite{Suzuki2016} and TbPtBi.~\cite{zhu.singh.20}

Time-reversal symmetry breaking introduced by the magnetic field may lead to the appearance of Weyl points.~\cite{cano.bradlyn.17} 
GdPtBi is an example of material in which this mechanism occurs.\cite{Hirschberger2016a}
Our results of the electronic structure calculations corroborate this scenario, six pairs of Weyl nodes exist in the vicinity of $\Gamma$ point, in accordance with the findings of Ref.\,\cite{Shekhar2018}.  
Furthermore, we also found that an increase in magnetic field strength leads to the gaping of Weyl nodes (see Fig.~\ref{fig.band}b, c). 
In general, the results of our electronic structure calculations are consistent with those previously reported for GdPtBi.\,\cite{Suzuki2016,Shekhar2018} 
While the Weyl points occur in proximity to the Fermi level, the electronic structure is only minimally impacted by magnetic field below the Fermi level. 
The density of states as a function of energy is shown in Fig.\,S3a (see Supplementary Materials) for various values of magnetic field. 
We therefore neglected this change when estimating the position of the Fermi level in the irradiated samples. 
By comparing the results of electronic structure calculations and carrier concentration obtained from Hall effect measurements, we estimated the position of Fermi level in pristine and electron irradiated samples (see Supplementary Materials for more details). 
In general, electron irradiation of GdPtBi caused quite small change in the Fermi level position. 
For pristine samples, the Fermi level is situated at $-30$\,meV, while for sample irradiated with maximal dose of $7$\,C$/$cm$^2$ Fermi level is located at $-130$\,meV, resulting in a relative shift of $100$\,meV, which is smaller than that we observed in LuPdBi.\cite{Ishihara2021a} 
The downward shift of Fermi level indicates that the Weyl nodes contribute less to the total magneto-transport properties. 
Consequently, we observed a smaller negative LMR with an increasing irradiation dose.

\begin{figure*}
	\includegraphics[width=\textwidth]{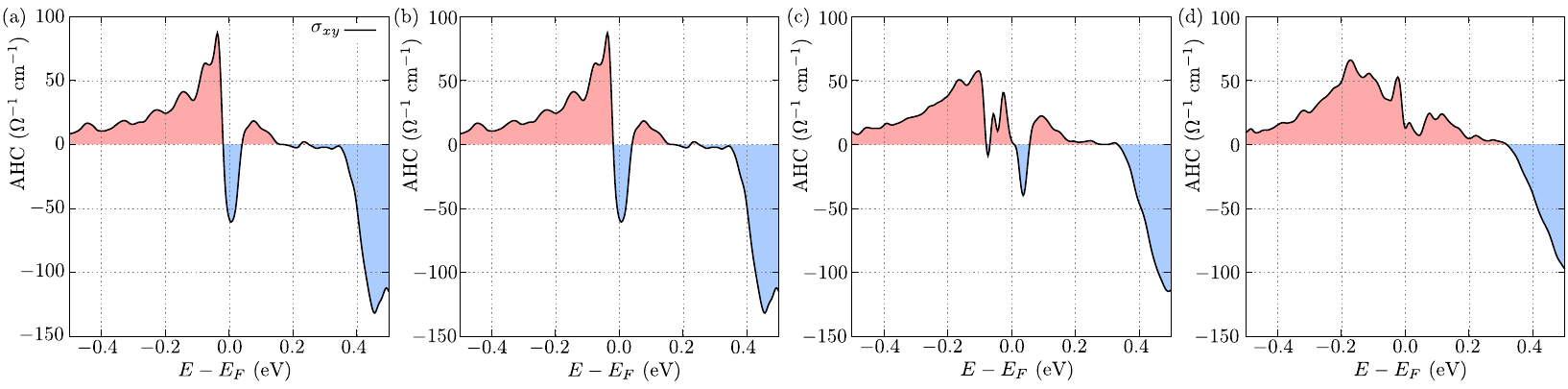}
	\caption{Theoretically obtained anomalous Hall conductivity as a function of energy in the presence of magnetic field of 2.5\,T (a), 5\,T (b), 7.5\,T (c) and 10\,T (d).
		\label{fig.theo_ahc}}
\end{figure*}

Additionally, the anomalous Hall conductivity (AHC) was calculated for several different values of magnetic field (Fig.~\ref{fig.theo_ahc}).
The AHC exhibits pronounced non-monotonic behaviour as a function of energy, and magnetic field exerts a significant influence on the AHC($E-E_F$) curves.  
There is a pronounced peak with the maximum close to 50\,meV below the Fermi level for 2.5 and 5\,T, but as magnetic field increases further the peak blurs and AHC becomes smaller. 
In general, the values of AHC obtained from theoretical calculations are of the same order of magnitude as obtained from Hall effect analysis. 
Moreover, both the theoretical and experimental data show complex variation of the anomalous Hall effect with respect to magnetic field and the position of the Fermi level. 
A discrepancy between two data sets may be attributed to the fact that in the calculations, the assumption of full polarization of magnetic moments was made.
This is in turn inconsistent with the values of magnetic field at which experiments were carried out, whereby GdPtBi is in the canted antiferromagnetic state. 
For GtPtBi, the magnetic moments are fully saturated above $\sim25$\,T.\cite{Shekhar2018}         
The realization of a large AHC around the Fermi level can be related to (i) the magnetic-filed-induced Weyl points,~\cite{burkov.14} and/or (ii) the avoided crossing of the electronic bands.~\cite{yao.kleinman.04}
In the case of GdPtBi, both effects can give rise to a significant Berry phase contribution to the AHE.~\cite{Suzuki2016} 
However, our findings indicate that electron irradiation of GdPtBi results in a downward shift of the Fermi level, causing the Weyl nodes to move further away from it. 
Hence, the AHE may be mainly dominated by avoided band crossing in GdPtBi, particularly in the case of irradiated samples. 

\section{Conclusions}

In summary, we conducted a comprehensive investigation of the magnetotransport properties of pristine and high-energy-electron-irradiated samples of the half-Heusler Weyl semimetal GdPtBi. 
Comparing the results of Hall effect measurements with theoretical calculations of the electronic structure, we concluded that irradiation shifts the Fermi level far from the magnetic field-induced Weyl nodes. 
This shift reduces the contribution of Weyl states to observed magnetotransport, evidenced by the decrease in negative longitudinal magnetoresistance as the irradiation dose increases. 
We observed that negative LMR is proportional to the carrier concentration in GdPtBi, while transverse MR is proportional to the squared carrier mobility.
Furthermore, we observed the anomalous Hall effect, which is strongly influenced by electron irradiation. 
The magnitude of AHE and the magnetic field values at which AHE peaks vary in a complex manner with irradiation dose change, consistent with theoretically calculated anomalous Hall conductivity. 
Our investigation shows that in GdPtBi impact of Weyl states on the magnetotransport properties is robust against the moderate Fermi level tuning. 
This finding is applicable to other half-Heusler Weyl semimetals, thereby opening new avenues for their study, as there are not many materials in which Weyl nodes lie sufficiently close to the Fermi level to still observe their contribution to magnetotransport properties.      

\section{Methods}

\subsection{Single crystals growth and characterization}
GdPtBi single crystals were grown from Bi-flux as described in Ref.~\cite{Pavlosiuk2016d}. 
X-ray powder diffraction experiment were performed on an X’pert Pro (PANalytical) diffractometer with CuK$_{\alpha}$ radiation. 
Refinement of the crystal structure was performed by the Rietveld method using Fullprof software,\cite{Rodriguez-Carvajal1993} its results confirmed the MgAgAs-type crystal structure of studied material, and the obtained lattice parameter is $a=6.684\,\rm{\AA}$, identical to that reported earlier.\cite{Suzuki2016} 
The high quality of synthesized single crystals was confirmed and their orientation was assessed by the Laue backscattering technique using Proto LAUE-COS system. 

\subsection{Electronic transport measurements}
Electrical resistivity, magnetoresistance and Hall effect were investigated using the Quantum Design PPMS platform. 
A standard four-probe ac-current technique was used and electrical contacts were prepared with a 50\,$\rm{\mu}$m-diameter silver wire attached to the samples with silver epoxy paste.

\subsection{Electron irradiation}
Irradiations by 2.5\,MeV electrons were performed at the SIRIUS platform operated Laboratoire des Solides Irradiés at École Polytechnique (Palaiseau, France). 
Electron beam was generated by NEC  Pelltron accelerator coupled with low temperature irradiation chamber. 
During irradiation, sample was immersed in liquid hydrogen at 22K, delivered by close cycle re-condenser. 
Irradiation dose was monitored by integration of current collected on Faraday cup placed behind sample. 
Irradiation area was limited by 5\,mm diameter diaphragm and uniformity of damage was granted by sweeping the beam vertically and horizontally at two non-commensurate frequencies. 
Beam current density was limited to 10\,$\mu\rm{A}/\rm{cm}^2$ to prevent heating of the sample. 
At electron energy used in our experiment, 2.5\,MeV electrons are relativistic and Rutherford collision are the main channel of energy transfer from impinging electrons to the atoms of the target. 
This process was examined by N.\,F.\,Mott\cite{Mott1929} and results, probabilities of ejection of the ions from it site and creation of vacancy – interstitial (Frenkel pair) were tabulated by Oen.\cite{Oen1973} 
Using those tables, we estimated effective cross-sections for defect creation on each sublattice (see Fig.\,S4 in Supplementary Materials). 
Removal of samples from irradiation chamber and transfer at room temperature to another instrument lead to partial annealing of defects, mainly interstitials characterized by lower migration energy. 
We assume that remaining damage is of vacancy type. 
Penetration range of 2.5\,MeV electrons is over 1\,mm, which another argument supporting uniform distribution of defects.
In total, we irradiated and then studied 5 samples (named as $\#1\!-\!\#5$), which were cut from the same single crystal. 
Samples $\#1$ and $\#2$ were also investigated before the irradiation.  

\subsection{Ab initio calculation}
The first-principles density functional theory (DFT) calculations were performed using the projector augmented-wave (PAW) potentials~\cite{blochl.94} implemented in 
the Vienna Ab-initio Simulation Package (VASP).~\cite{kresse.hafner.94,kresse.furthmuller.96,kresse.joubert.99}
The calculations comprising the spin-orbit coupling (SOC) were performed with the generalized gradient approximation (GGA) under the modified Becke-Johnson (mBJ) exchanged potential.~\cite{becke.johnson.06,tran.blaha.09,camargo.baquero.12}
The energy cutoff for the plane-wave expansion was set to $350$~eV.
The electronic properties were calculated using $10\!\times\!10\!\times 10$ $\Gamma$-centered {\bf k}-point grid within the Monkhorst-Pack scheme.~\cite{monkhorst.pack.76}
As a convergence criterion we took the energy change below $10^{-10}$~eV for electronic degrees of freedom.
We assumed the experiential lattice constant, while the Gd $f-$orbitals were treated as core states.
The band structure from the exact DFT calculations, was used to construct the tight binding model (TBM) using Wannier90 software.~\cite{pizzi.vitale.20}
We started the calculations from $s$, $p$, $d$ orbitals of all atoms, which gave TBM with $54$ orbitals and $108$ bands.
The band structure in the presence of magnetic field was calculated from the TBM with additional Zeeman term.

\section*{Author contributions}
Orest Pavlosiuk: conceptualization, data curation, formal analysis, investigation, methodology, visualization, writing – original draft, writing – review \& editing. 
Piotr Wi{\'{s}}niewski: conceptualization, formal analysis, methodology, writing – review \& editing.
Romain Grasset: investigation, methodology, visualization, writing – review \& editing.
Marcin Konczykowski: conceptualization, data curation, formal analysis, funding acquisition, investigation, methodology, writing – review \& editing.
Andrzej Ptok: data curation, formal analysis, investigation, visualization, writing – original draft, writing – review \& editing. 
Dariusz Kaczorowski: conceptualization, funding acquisition, methodology, project administration, writing – review \& editing.

\section*{Acknowledgement}

This research was financially supported by the National Science Centre of Poland, grant no. 2021/40/Q/ST5/00066; and by the Agence Nationale de la Recherche project “DYNTOP” ANR-22-CE30-0026-01.
The authors acknowledge support from the EMIR\&A French network (FR CNRS 3618) on the platform SIRIUS.


\newpage

\pagenumbering{arabic}
\setcounter{page}{1}
\renewcommand{\thepage}{S\arabic{page}}
\setcounter{figure}{0}
\renewcommand{\thefigure}{S\arabic{figure}}
\renewcommand{\thetable}{S\arabic{table}}
\renewcommand{\theequation}{S\arabic{equation}}
\newcommand{\cor}[1]{\color{red}{#1}}
\newcommand{\cob}[1]{\color{blue}{#1}}
\newcommand{\OP}[1]{\color{teal}{#1}}
\renewcommand*{\familydefault}{\sfdefault}
\bibliographystyle{naturemag}
\renewcommand{\citenumfont}[1]{S#1}
\renewcommand{\bibnumfmt}[1]{[S#1]}

	\begin{large}{\bf Supplementary Materials for} \end{large}\\\\
	\begin{Large}{\bf Tuning of anomalous magnetotransport properties in half-Heusler topological semimetal GdPtBi}\end{Large}\\\\
	\begin{large}Orest Pavlosiuk$^{1,*}$, Piotr Wi\'{s}niewski$^1$, Romain Grasset$^2$, Marcin Konczykowski$^2$, Andrzej Ptok$^3$, Dariusz Kaczorowski$^{1,*}$
	\end{large}\\\\
	{\it $^1$~Institute of Low Temperature and Structure Research, Polish Academy of Sciences, ul.~Ok{\'{o}}lna~2, 50-422 Wroc{\l}aw, Poland\\
		$^2$~Laboratoire des Solides Irradiés, École Polytechnique, 91128 Palaiseau, France\\
		$^3$~Institute of Nuclear Physics, Polish Academy of Sciences,
		W. E. Radzikowskiego 152, PL-31342 Krak\'{o}w, Poland}\\

	\subsection*{Magnetoresistance of pristine samples}
	\begin{figure}[h]
		\includegraphics[width=8cm]{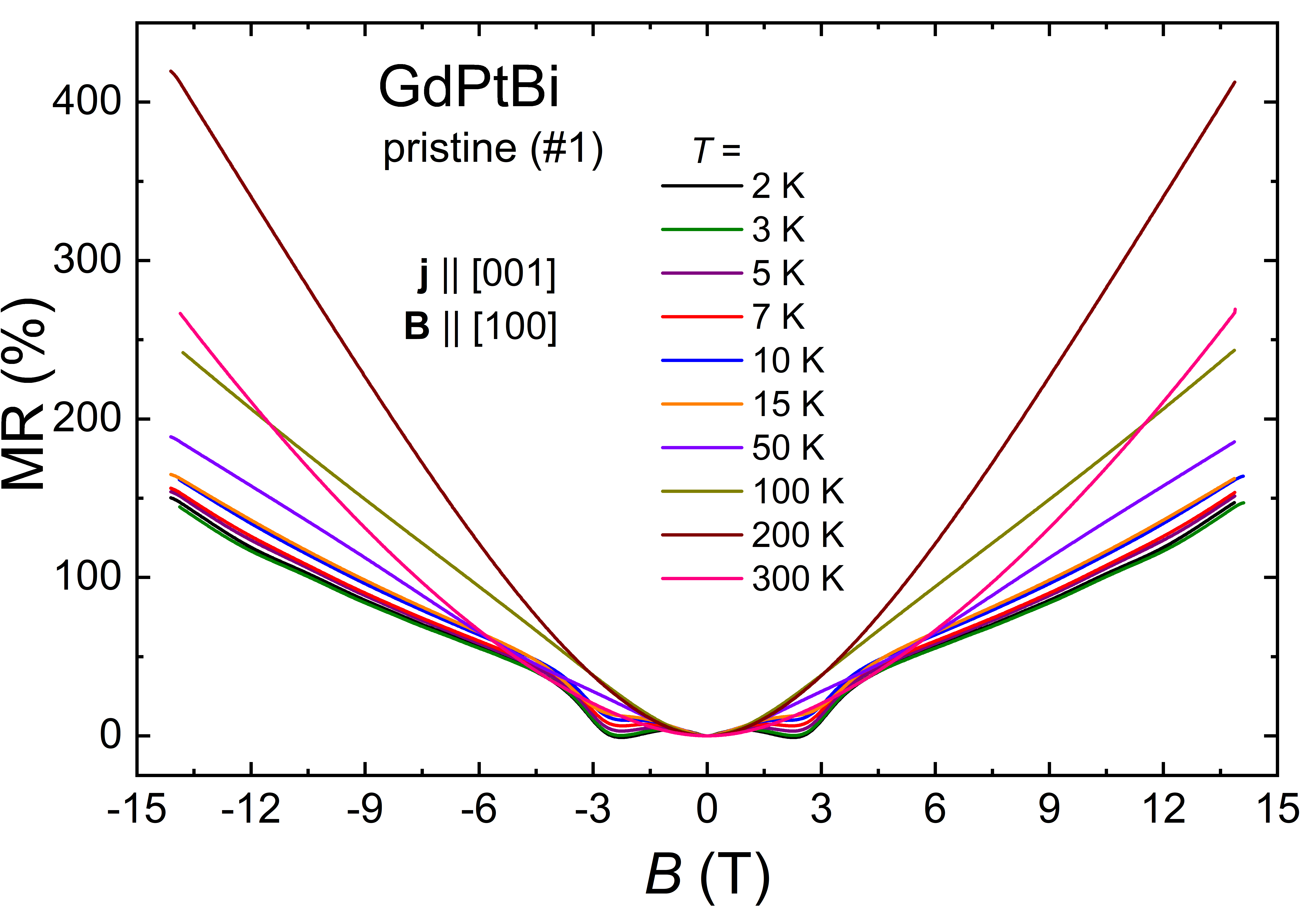}
		\includegraphics[width=8cm]{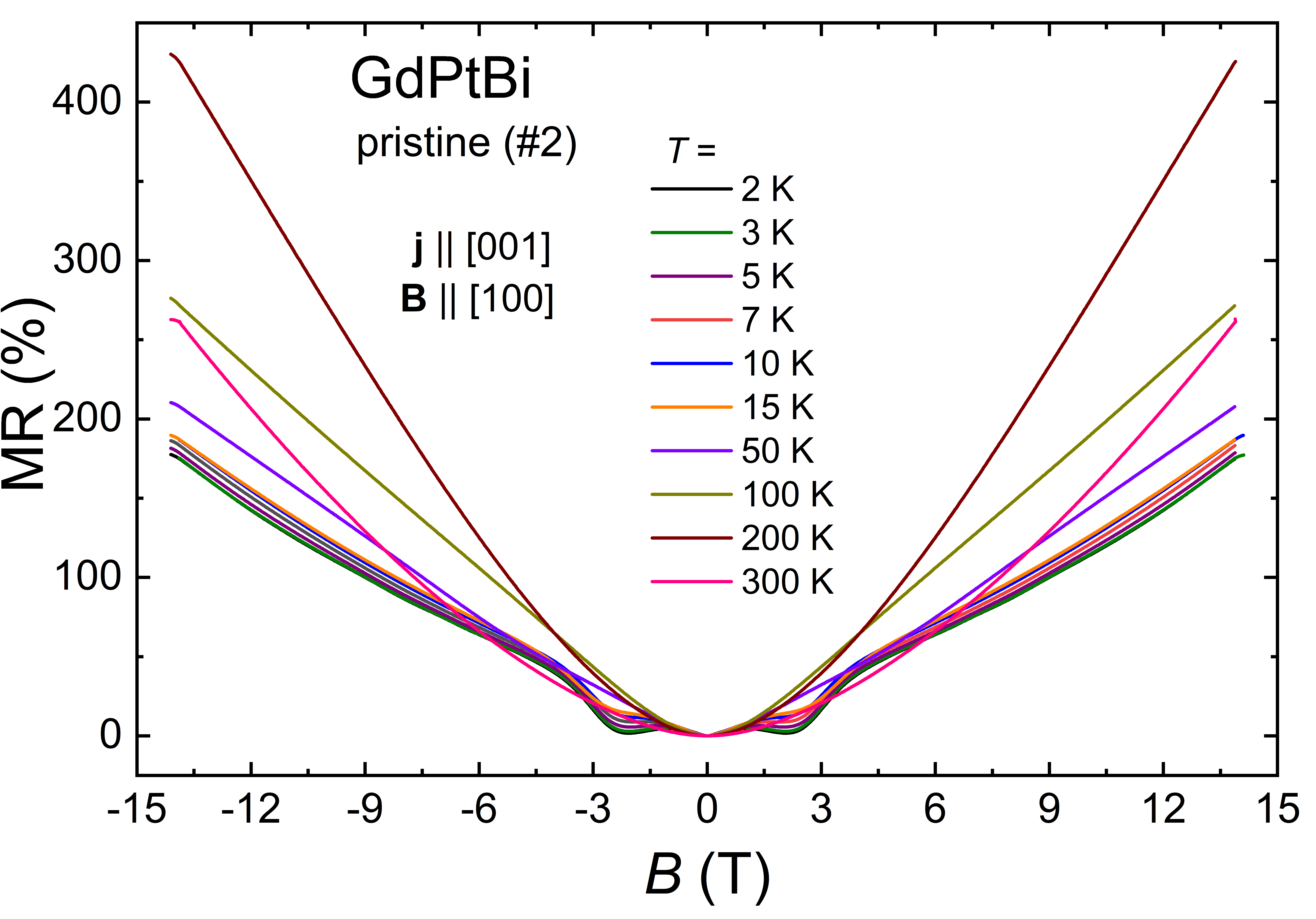}
		\caption{Magnetoresistance isotherms as a function of magnetic field measured at several different temperatures for pristine sample $\#1$ (a) and pristine sample $\#2$ (b).      
			\label{MR_fig}}
	\end{figure}
	\newpage
	\subsection*{Separation of anomalous Hall contribution}
	\begin{figure}[h]
		\includegraphics[width=16cm]{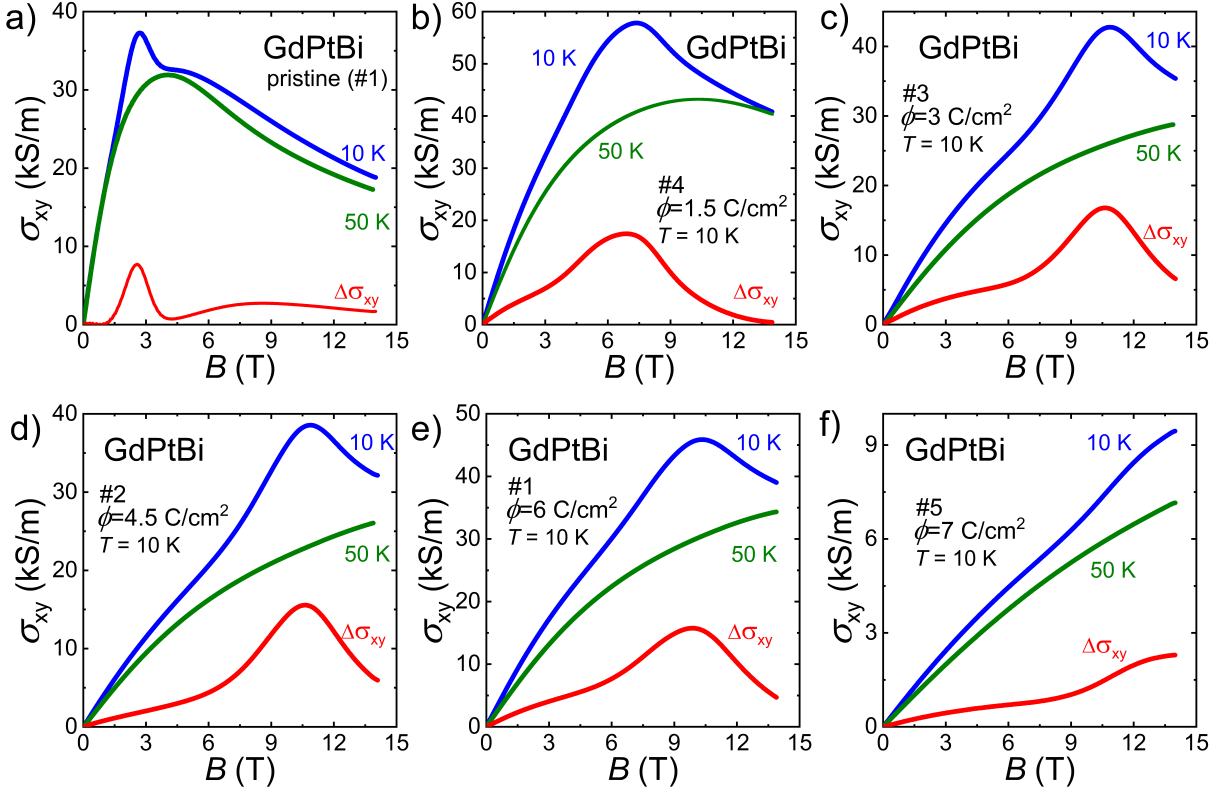}
		\caption{Magnetic field dependence of Hall conductivity measured at $T=10$\,K (blue line), at $T=50$\,K (olive line) and their difference (red line) for pristine sample $\#1$ (a), sample irradiated with $\phi=1.5$\,C/cm$^2$ (b), $\phi=3$\,C/cm$^2$ (c), $\phi=4.5$\,C/cm$^2$ (d), $\phi=6$\,C/cm$^2$ (e) and $\phi=7$\,C/cm$^2$\,(f).          
			\label{AHE_separ}}
	\end{figure}

	The anomalous Hall effect has been observed in a number of half-Heusler compounds.\cite{Zhu2023b_1,Shekhar2018_1,Suzuki2016_1,Pavlosiuk2020_1,zhu.singh.20_1,Chen2021d1,Singha2019a} 
	There are several different approaches that can be taken in order to extract the contribution of AHE to the total Hall effect observed in $RE$PtBi half-Heusler compounds.\cite{Shekhar2018_1,Suzuki2016_1,Pavlosiuk2020_1,zhu.singh.20_1} 
	It is evident that $\sigma_{xy}(B)$ curves obtained at $T=10$\,K for all studied samples, show an anomaly due to the AHE. 
	Furthermore, this anomaly changes its position with respect to the magnetic field scale (see Fig.\,\ref{AHE_separ}). 
	In pristine samples, AHE reaches its maximum in the low magnetic field region. 
	In contrast, in irradiated samples, AHE shifts towards a stronger magnetic field. 
	Accordingly, methods based on the fitting of Drude model,\cite{Pavlosiuk2020_1,Suzuki2016_1} cannot be applied in our case.
	Therefore, we used the approach previously reported for AHE analysis in TbPtBi\cite{zhu.singh.20_1}, which is the most suitable to our data set. 
	In this approach, it has been assumed that at sufficiently high temperatures, AHE can be completely neglected, and the total Hall signal at this particular temperature can be considered to be equal to the ordinary Hall effect contribution at low temperature. 
	A comparison of $\sigma_{xy}(B)$ isotherms measured in the temperature range 2-300\,K revealed that the lowest temperature at which AHE is negligible is 50\,K. 
	Furthermore, there is a slight difference between $\rho_{xy}(B)$ measured at $T=10$\,K and $T=50$\,K, therefore it is a good assumption that $\sigma_{xy}(B)$ at $T=50$\,K corresponds to the ordinary Hall contribution at $T=10$\,K. 
	Fig.\,\ref{AHE_separ}a-f, shows the magnetic field dependence of $\sigma_{xy}$ at $T=10$\,K (blue curves) and $T=50$\,K (olive curves), as well as their difference, which corresponds to AHE ($\sigma^A_{xy}=\sigma_{xy}(10\,K)-\sigma_{xy}(50\,K)$, red curves) for particular samples.     
	
	\subsection*{Determination of the Fermi level position in the studied samples}

	\begin{figure}[h]
		\includegraphics[width=8cm]{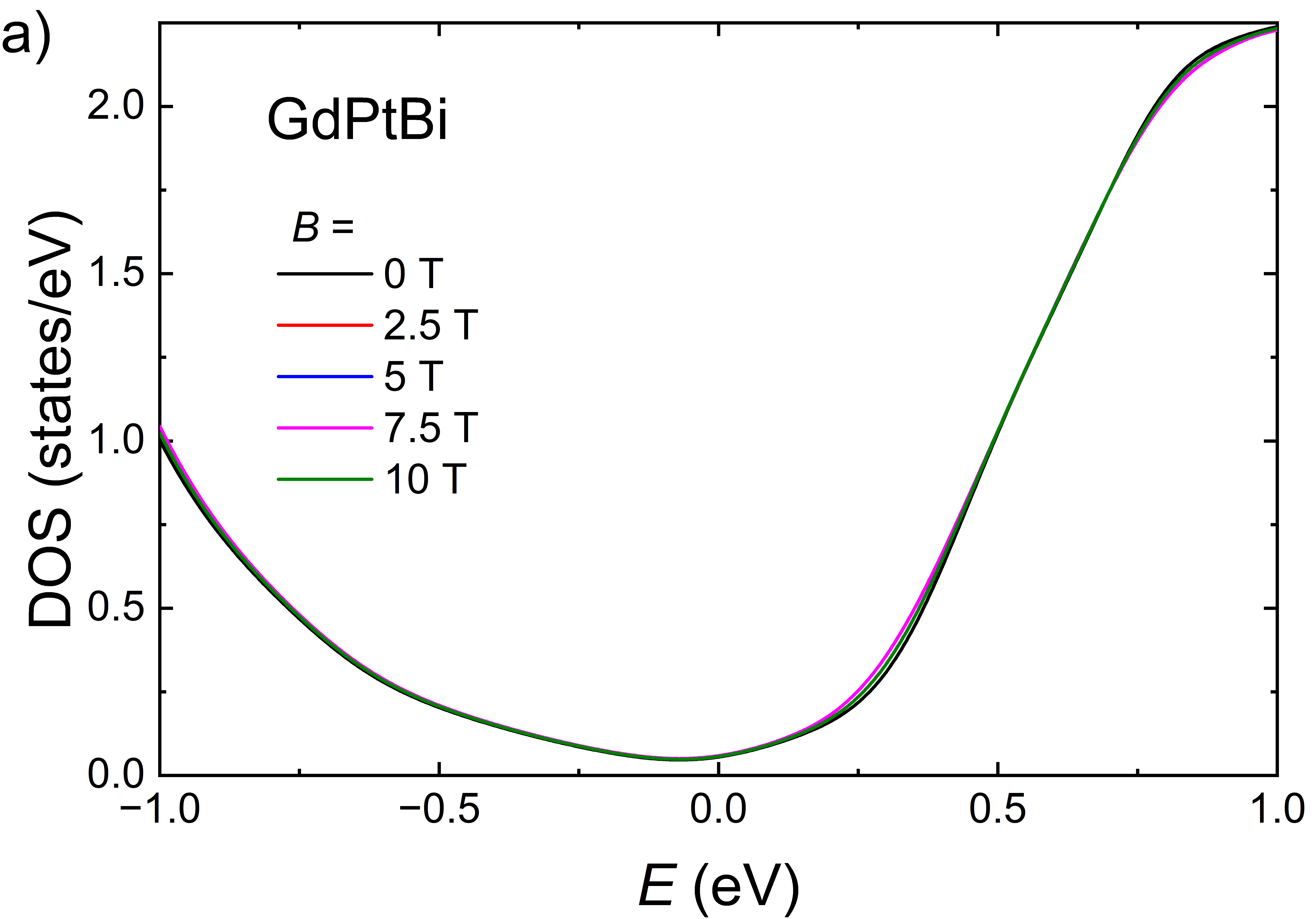}
		\includegraphics[width=6.5cm]{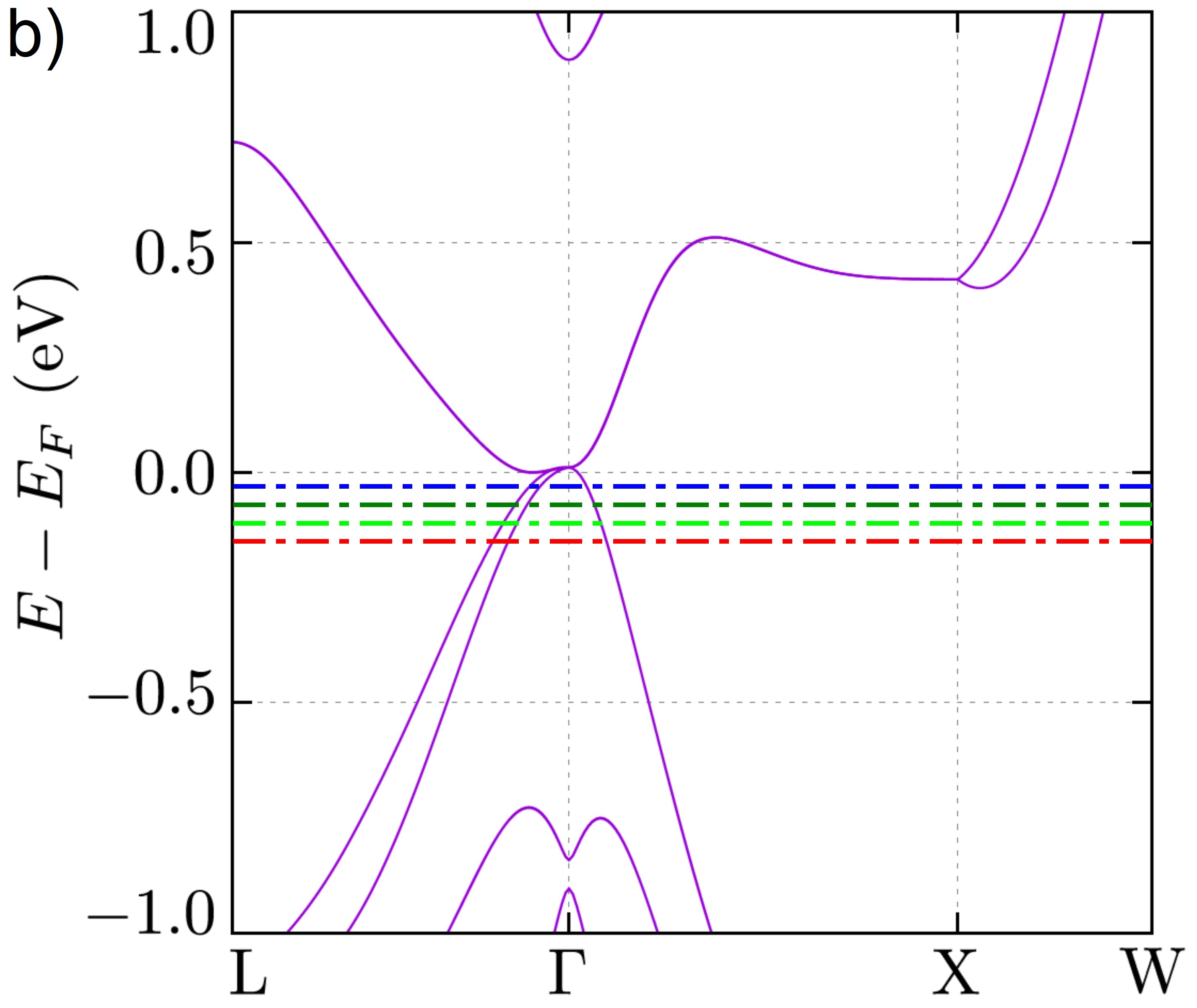}
		\caption{(a) Density of states as a function of energy for several different values of magnetic field; (b) Calculated electronic band structure in the absence of magnetic field. The horizontal blue, olive, green and red dashed lines correspond to Fermi levels obtained from the Hall carrier concentrations in the pristine, 1.5, 3, and 7\,C/cm$^2$ irradiated samples, respectively.
			\label{n_vs_E}}
	\end{figure}

	There is slight difference in the density of states calculated for zero magnetic field and for several different values of the magnetic field in the range 2.5-10\,T (see Fig.\ref{n_vs_E}a). 
	Therefore, we used the results of electronic structure calculations obtained for zero magnetic field to determine the position of the Fermi level in the studied samples. 
	Using the value of the Fermi wave vector $k_F$ calculated for $\Gamma-X$ high symmetry line for different values of energies, we calculated the carrier concentration using the formula $n=V_F/(4\pi^3)$, where $V_F$ is volume of the Fermi pocket, which we assumed to be a sphere.
	By comparing the theoretically calculated carrier concentrations with those determined experimentally from the Hall effect data, we estimated the position of Fermi level in pristine and irradiated samples, as depicted schematically in Fig.\,\ref{n_vs_E}b.
	
	\subsection*{Calculated cross-sections for Frenkel pair production}

	\begin{figure}[h]
		\includegraphics[width=8cm]{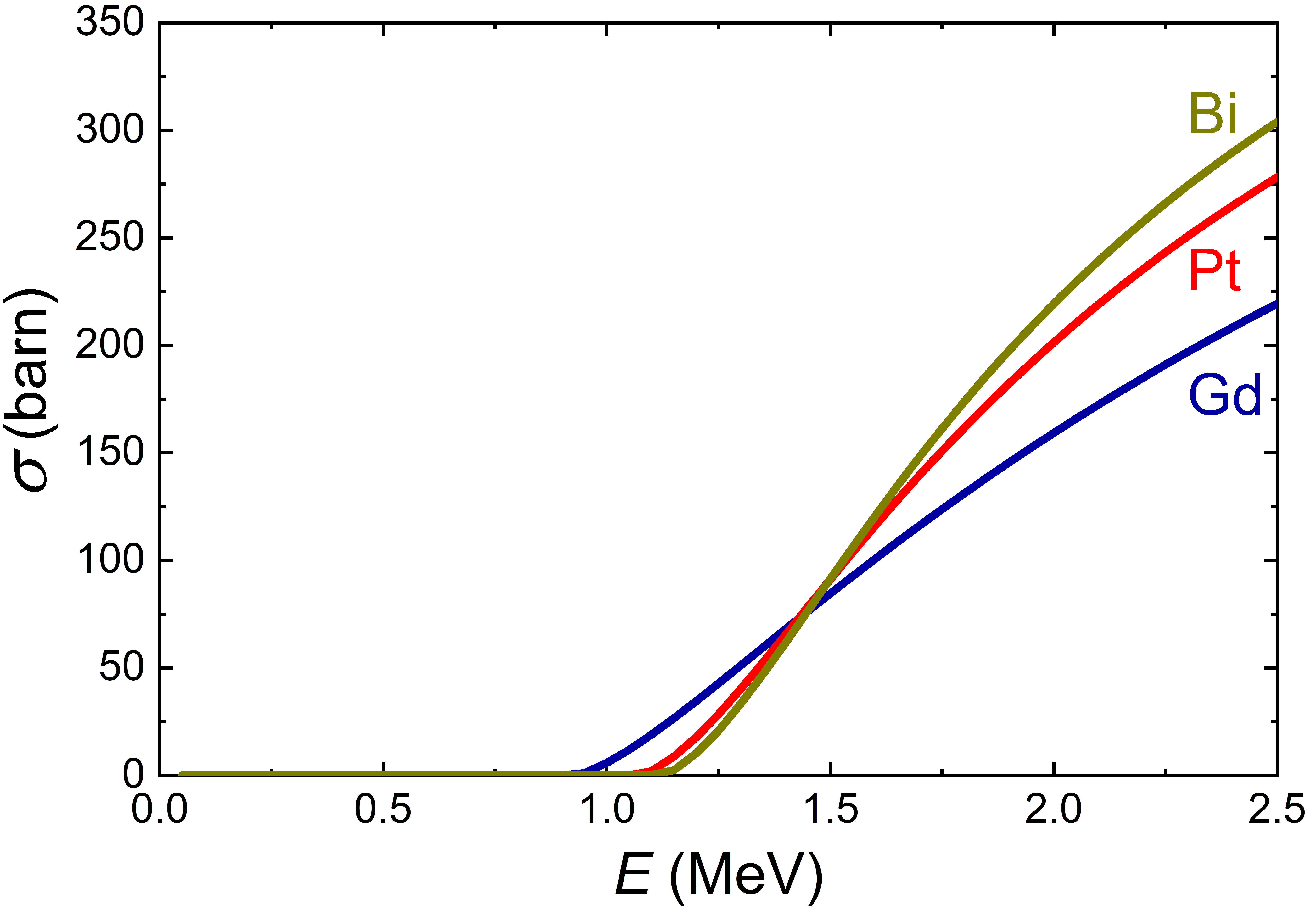}
		\caption{Frenkel pair production cross-sections ($\sigma$) for Gd, Pt and Bi sublattices as a function of electron energy ($E$).
			\label{Frenkel}}
	\end{figure}
	%
	

\begin{thebibliography}{10}
	\expandafter\ifx\csname url\endcsname\relax
	\def\url#1{\texttt{#1}}\fi
	\expandafter\ifx\csname urlprefix\endcsname\relax\def\urlprefix{URL }\fi
	\providecommand{\bibinfo}[2]{#2}
	\providecommand{\eprint}[2][]{\url{#2}}
	
	\bibitem{Hasan2010a}
	\bibinfo{author}{Hasan, M.~Z.} \& \bibinfo{author}{Kane, C.~L.}
	\newblock \bibinfo{title}{{Colloquium: Topological insulators}}.
	\newblock \emph{\bibinfo{journal}{Rev. Mod. Phys.}}
	\textbf{\bibinfo{volume}{82}}, \bibinfo{pages}{3045} (\bibinfo{year}{2010}).
	\newblock \urlprefix\url{http://doi.org/10.1103/RevModPhys.82.3045}.
	
	\bibitem{Armitage2018}
	\bibinfo{author}{Armitage, N.~P.}, \bibinfo{author}{Mele, E.~J.} \&
	\bibinfo{author}{Vishwanath, A.}
	\newblock \bibinfo{title}{{Weyl and Dirac semimetals in three-dimensional
			solids}}.
	\newblock \emph{\bibinfo{journal}{Rev. Mod. Phys.}}
	\textbf{\bibinfo{volume}{90}}, \bibinfo{pages}{015001}
	(\bibinfo{year}{2018}).
	\newblock \urlprefix\url{http://doi.org/10.1103/RevModPhys.90.015001}.
	
	\bibitem{Hasan2017}
	\bibinfo{author}{Hasan, M.~Z.}, \bibinfo{author}{Xu, S.-Y.},
	\bibinfo{author}{Belopolski, I.} \& \bibinfo{author}{Huang, S.-M.}
	\newblock \bibinfo{title}{{Discovery of Weyl Fermion Semimetals and Topological
			Fermi Arc States}}.
	\newblock \emph{\bibinfo{journal}{Annu. Rev. Condens. Matter Phys.}}
	\textbf{\bibinfo{volume}{8}}, \bibinfo{pages}{289} (\bibinfo{year}{2017}).
	\newblock
	\urlprefix\url{https://doi.org/10.1146/annurev-conmatphys-031016-025225}.
	
	\bibitem{Hellenbrandt2004}
	\bibinfo{author}{Hellenbrandt, M.}
	\newblock \bibinfo{title}{{The inorganic crystal structure database (ICSD) -
			Present and future}}.
	\newblock \emph{\bibinfo{journal}{Crystallogr. Rev.}}
	\textbf{\bibinfo{volume}{10}}, \bibinfo{pages}{17} (\bibinfo{year}{2004}).
	\newblock \urlprefix\url{https://doi.org/10.1080/08893110410001664882}.
	
	\bibitem{Zhang2019g}
	\bibinfo{author}{Zhang, T.} \emph{et~al.}
	\newblock \bibinfo{title}{{Catalogue of topological electronic materials}}.
	\newblock \emph{\bibinfo{journal}{Nature}} \textbf{\bibinfo{volume}{566}},
	\bibinfo{pages}{475} (\bibinfo{year}{2019}).
	\newblock \urlprefix\url{https://doi.org/10.1038/s41586-019-0944-6}.
	
	\bibitem{Vergniory2019}
	\bibinfo{author}{Vergniory, M.~G.} \emph{et~al.}
	\newblock \bibinfo{title}{{A complete catalogue of high-quality topological
			materials}}.
	\newblock \emph{\bibinfo{journal}{Nature}} \textbf{\bibinfo{volume}{566}},
	\bibinfo{pages}{480} (\bibinfo{year}{2019}).
	\newblock \urlprefix\url{http://doi.org/10.1038/s41586-019-0954-4}.
	
	\bibitem{Tang2019a}
	\bibinfo{author}{Tang, F.}, \bibinfo{author}{Po, H.~C.},
	\bibinfo{author}{Vishwanath, A.} \& \bibinfo{author}{Wan, X.}
	\newblock \bibinfo{title}{{Comprehensive search for topological materials using
			symmetry indicators}}.
	\newblock \emph{\bibinfo{journal}{Nature}} \textbf{\bibinfo{volume}{566}},
	\bibinfo{pages}{486} (\bibinfo{year}{2019}).
	\newblock \urlprefix\url{http://doi.org/10.1038/s41586-019-0937-5}.
	
	\bibitem{Mun2022}
	\bibinfo{author}{Mun, E.} \& \bibinfo{author}{Bud'ko, S.~L.}
	\newblock \bibinfo{title}{{RPtBi: Magnetism and topology}}.
	\newblock \emph{\bibinfo{journal}{MRS Bull.}} \textbf{\bibinfo{volume}{47}},
	\bibinfo{pages}{609} (\bibinfo{year}{2022}).
	\newblock \urlprefix\url{https://doi.org/10.1557/s43577-022-00353-y}.
	
	\bibitem{Hirschberger2016a}
	\bibinfo{author}{Hirschberger, M.} \emph{et~al.}
	\newblock \bibinfo{title}{{The chiral anomaly and thermopower of Weyl fermions
			in the half-Heusler GdPtBi}}.
	\newblock \emph{\bibinfo{journal}{Nat. Mater.}} \textbf{\bibinfo{volume}{15}},
	\bibinfo{pages}{1161} (\bibinfo{year}{2016}).
	\newblock \urlprefix\url{https://doi.org/10.1038/nmat4684}.
	
	\bibitem{Shekhar2018}
	\bibinfo{author}{Shekhar, C.} \emph{et~al.}
	\newblock \bibinfo{title}{{Anomalous Hall effect in Weyl semimetal half-Heusler
			compounds RPtBi (R = Gd and Nd)}}.
	\newblock \emph{\bibinfo{journal}{Proc. Natl. Acad. Sci. U.S.A.}}
	\textbf{\bibinfo{volume}{115}}, \bibinfo{pages}{9140} (\bibinfo{year}{2018}).
	\newblock \urlprefix\url{https://doi.org/10.1073/pnas.1810842115}.
	
	\bibitem{Kumar2018}
	\bibinfo{author}{Kumar, N.}, \bibinfo{author}{Guin, S.~N.},
	\bibinfo{author}{Felser, C.} \& \bibinfo{author}{Shekhar, C.}
	\newblock \bibinfo{title}{{Planar Hall effect in the Weyl semimetal GdPtBi}}.
	\newblock \emph{\bibinfo{journal}{Phys. Rev. B}} \textbf{\bibinfo{volume}{98}},
	\bibinfo{pages}{041103} (\bibinfo{year}{2018}).
	\newblock \urlprefix\url{https://doi.org/10.1103/PhysRevB.98.041103}.
	
	\bibitem{Suzuki2016}
	\bibinfo{author}{Suzuki, T.} \emph{et~al.}
	\newblock \bibinfo{title}{{Large anomalous Hall effect in a half-Heusler
			antiferromagnet}}.
	\newblock \emph{\bibinfo{journal}{Nat. Phys.}} \textbf{\bibinfo{volume}{12}},
	\bibinfo{pages}{1119} (\bibinfo{year}{2016}).
	\newblock \urlprefix\url{https://doi.org/10.1038/nphys3831}.
	
	\bibitem{Pavlosiuk2019}
	\bibinfo{author}{Pavlosiuk, O.}, \bibinfo{author}{Kaczorowski, D.} \&
	\bibinfo{author}{Wi{\'{s}}niewski, P.}
	\newblock \bibinfo{title}{{Negative longitudinal magnetoresistance as a sign of
			a possible chiral magnetic anomaly in the half-Heusler antiferromagnet
			DyPdBi}}.
	\newblock \emph{\bibinfo{journal}{Phys. Rev. B}} \textbf{\bibinfo{volume}{99}},
	\bibinfo{pages}{125142} (\bibinfo{year}{2019}).
	\newblock \urlprefix\url{https://doi.org/10.1103/PhysRevB.99.125142}.
	
	\bibitem{Pavlosiuk2020}
	\bibinfo{author}{Pavlosiuk, O.}, \bibinfo{author}{Fa{\l}at, P.},
	\bibinfo{author}{Kaczorowski, D.} \& \bibinfo{author}{Wi{\'{s}}niewski, P.}
	\newblock \bibinfo{title}{{Anomalous Hall effect and negative longitudinal
			magnetoresistance in half-Heusler topological semimetal candidates TbPtBi and
			HoPtBi}}.
	\newblock \emph{\bibinfo{journal}{APL Mater.}} \textbf{\bibinfo{volume}{8}},
	\bibinfo{pages}{111107} (\bibinfo{year}{2020}).
	\newblock \urlprefix\url{https://doi.org/10.1063/5.0026956}.
	
	\bibitem{Chen2020a}
	\bibinfo{author}{Chen, J.} \emph{et~al.}
	\newblock \bibinfo{title}{{Chiral-anomaly induced large negative
			magnetoresistance and nontrivial $\pi$-Berry phase in half-Heusler compounds
			RPtBi (R=Tb, Ho, and Er)}}.
	\newblock \emph{\bibinfo{journal}{Appl. Phys. Lett.}}
	\textbf{\bibinfo{volume}{116}}, \bibinfo{pages}{222403}
	(\bibinfo{year}{2020}).
	\newblock \urlprefix\url{http://doi.org/10.1063/5.0007528}.
	
	\bibitem{Guo2018}
	\bibinfo{author}{Guo, C.~Y.} \emph{et~al.}
	\newblock \bibinfo{title}{{Evidence for Weyl fermions in a canonical
			heavy-fermion semimetal YbPtBi}}.
	\newblock \emph{\bibinfo{journal}{Nat. Commun.}} \textbf{\bibinfo{volume}{9}},
	\bibinfo{pages}{4622} (\bibinfo{year}{2018}).
	\newblock \urlprefix\url{https://doi.org/10.1038/s41467-018-06782-1}.
	
	\bibitem{Chen2021h}
	\bibinfo{author}{Chen, J.} \emph{et~al.}
	\newblock \bibinfo{title}{{Unconventional Anomalous Hall Effect in the Canted
			Antiferromagnetic Half‐Heusler Compound DyPtBi}}.
	\newblock \emph{\bibinfo{journal}{Adv. Funct. Mater.}} \bibinfo{pages}{2107526}
	(\bibinfo{year}{2021}).
	\newblock \urlprefix\url{https://doi.org/10.1002/adfm.202107526}.
	
	\bibitem{Son2013}
	\bibinfo{author}{Son, D.~T.} \& \bibinfo{author}{Spivak, B.~Z.}
	\newblock \bibinfo{title}{{Chiral anomaly and classical negative
			magnetoresistance of Weyl metals}}.
	\newblock \emph{\bibinfo{journal}{Phys. Rev. B}} \textbf{\bibinfo{volume}{88}},
	\bibinfo{pages}{104412} (\bibinfo{year}{2013}).
	\newblock \urlprefix\url{https://doi.org/10.1103/PhysRevB.88.104412}.
	
	\bibitem{Zhu2023b}
	\bibinfo{author}{Zhu, Y.} \emph{et~al.}
	\newblock \bibinfo{title}{{Large anomalous Hall effect and negative
			magnetoresistance in half-topological semimetals}}.
	\newblock \emph{\bibinfo{journal}{Commun. Phys.}} \textbf{\bibinfo{volume}{6}},
	\bibinfo{pages}{346} (\bibinfo{year}{2023}).
	\newblock \urlprefix\url{https://doi.org/10.1038/s42005-023-01469-6}.
	
	\bibitem{Sun2021}
	\bibinfo{author}{Sun, Z.~L.} \emph{et~al.}
	\newblock \bibinfo{title}{{Pressure-controlled anomalous Hall conductivity in
			the half-Heusler antiferromagnet GdPtBi}}.
	\newblock \emph{\bibinfo{journal}{Phys. Rev. B}}
	\textbf{\bibinfo{volume}{103}}, \bibinfo{pages}{085116}
	(\bibinfo{year}{2021}).
	\newblock \urlprefix\url{https://doi.org/10.1103/PhysRevB.103.085116}.
	
	\bibitem{Ishihara2021a}
	\bibinfo{author}{Ishihara, K.} \emph{et~al.}
	\newblock \bibinfo{title}{{Tuning the Parity Mixing of Singlet-Septet Pairing
			in a Half-Heusler Superconductor}}.
	\newblock \emph{\bibinfo{journal}{Phys. Rev. X}} \textbf{\bibinfo{volume}{11}},
	\bibinfo{pages}{041048} (\bibinfo{year}{2021}).
	\newblock \urlprefix\url{https://doi.org/10.1103/PhysRevX.11.041048}.
	
	\bibitem{Sitnicka2023}
	\bibinfo{author}{Sitnicka, J.} \emph{et~al.}
	\newblock \bibinfo{title}{{Fermi-level dependence of magnetism and
			magnetotransport in the magnetic topological insulators Bi$_2$Te$_3$ and
			BiSbTe$_3$ containing self-organized MnBi$_2$Te$_4$ septuple layers}}.
	\newblock \emph{\bibinfo{journal}{Phys. Rev. B}}
	\textbf{\bibinfo{volume}{107}}, \bibinfo{pages}{214424}
	(\bibinfo{year}{2023}).
	\newblock \urlprefix\url{https://doi.org/10.1103/PhysRevB.107.214424}.
	
	\bibitem{Mun2016a}
	\bibinfo{author}{Mun, E.}, \bibinfo{author}{Bud'ko, S.~L.} \&
	\bibinfo{author}{Canfield, P.~C.}
	\newblock \bibinfo{title}{{Robust tunability of magnetoresistance in
			half-Heusler RPtBi (R=Gd, Dy, Tm, and Lu) compounds}}.
	\newblock \emph{\bibinfo{journal}{Phys. Rev. B}} \textbf{\bibinfo{volume}{93}},
	\bibinfo{pages}{115134} (\bibinfo{year}{2016}).
	\newblock \urlprefix\url{http://doi.org/10.1103/PhysRevB.93.115134}.
	\newblock \eprint{1602.01194}.
	
	\bibitem{Mizukami2014}
	\bibinfo{author}{Mizukami, Y.} \emph{et~al.}
	\newblock \bibinfo{title}{{Disorder-induced topological change of the
			superconducting gap structure in iron pnictides}}.
	\newblock \emph{\bibinfo{journal}{Nat. Commun.}} \textbf{\bibinfo{volume}{5}},
	\bibinfo{pages}{5657} (\bibinfo{year}{2014}).
	\newblock \urlprefix\url{https://doi.org/10.1038/ncomms6657}.
	
	\bibitem{Pavlosiuk2016b}
	\bibinfo{author}{Pavlosiuk, O.}, \bibinfo{author}{Kaczorowski, D.} \&
	\bibinfo{author}{Wi{\'{s}}niewski, P.}
	\newblock \bibinfo{title}{{Superconductivity and Shubnikov–de Haas
			oscillations in the noncentrosymmetric half-Heusler compound YPtBi}}.
	\newblock \emph{\bibinfo{journal}{Phys. Rev. B}} \textbf{\bibinfo{volume}{94}},
	\bibinfo{pages}{035130} (\bibinfo{year}{2016}).
	\newblock \urlprefix\url{https://doi.org/10.1103/PhysRevB.94.035130}.
	
	\bibitem{Zhang2020k}
	\bibinfo{author}{Zhang, H.} \emph{et~al.}
	\newblock \bibinfo{title}{{Field-induced magnetic phase transitions and the
			resultant giant anomalous Hall effect in the antiferromagnetic half-Heusler
			compound DyPtBi}}.
	\newblock \emph{\bibinfo{journal}{Phys. Rev. B}}
	\textbf{\bibinfo{volume}{102}}, \bibinfo{pages}{094424}
	(\bibinfo{year}{2020}).
	\newblock \urlprefix\url{https://doi.org/10.1103/PhysRevB.102.094424}.
	
	\bibitem{Liang2018a}
	\bibinfo{author}{Liang, S.} \emph{et~al.}
	\newblock \bibinfo{title}{{Experimental Tests of the Chiral Anomaly
			Magnetoresistance in the Dirac-Weyl Semimetals Na$_3$Bi and GdPtBi}}.
	\newblock \emph{\bibinfo{journal}{Phys. Rev. X}} \textbf{\bibinfo{volume}{8}},
	\bibinfo{pages}{031002} (\bibinfo{year}{2018}).
	\newblock \urlprefix\url{https://doi.org/10.1103/PhysRevX.8.031002}.
	
	\bibitem{Huang2015}
	\bibinfo{author}{Huang, X.} \emph{et~al.}
	\newblock \bibinfo{title}{{Observation of the Chiral-Anomaly-Induced Negative
			Magnetoresistance in 3D Weyl Semimetal TaAs}}.
	\newblock \emph{\bibinfo{journal}{Phys. Rev. X}} \textbf{\bibinfo{volume}{5}},
	\bibinfo{pages}{031023} (\bibinfo{year}{2015}).
	\newblock \urlprefix\url{https://doi.org/10.1103/PhysRevX.5.031023}.
	
	\bibitem{Hikami1980}
	\bibinfo{author}{Hikami, S.}, \bibinfo{author}{Larkin, A.} \&
	\bibinfo{author}{Nagaoka, Y.}
	\newblock \bibinfo{title}{{Spin-orbit interaction and magnetoresistance in the
			two dimensional random system}}.
	\newblock \emph{\bibinfo{journal}{Prog. Theor. Phys}}
	\textbf{\bibinfo{volume}{63}}, \bibinfo{pages}{707} (\bibinfo{year}{1980}).
	\newblock \urlprefix\url{https://doi.org/10.1143/PTP.63.707}.
	
	\bibitem{Pal2010}
	\bibinfo{author}{Pal, H.~K.} \& \bibinfo{author}{Maslov, D.~L.}
	\newblock \bibinfo{title}{{Necessary and sufficient condition for longitudinal
			magnetoresistance}}.
	\newblock \emph{\bibinfo{journal}{Phys. Rev. B}} \textbf{\bibinfo{volume}{81}},
	\bibinfo{pages}{214438} (\bibinfo{year}{2010}).
	\newblock \urlprefix\url{https://doi.org/10.1103/PhysRevB.81.214438}.
	
	\bibitem{Pavlosiuk2021}
	\bibinfo{author}{Pavlosiuk, O.}, \bibinfo{author}{Jezierski, A.},
	\bibinfo{author}{Kaczorowski, D.} \& \bibinfo{author}{Wi{\'{s}}niewski, P.}
	\newblock \bibinfo{title}{{Magnetotransport signatures of chiral magnetic
			anomaly in the half-Heusler phase ScPtBi}}.
	\newblock \emph{\bibinfo{journal}{Phys. Rev. B}}
	\textbf{\bibinfo{volume}{103}}, \bibinfo{pages}{205127}
	(\bibinfo{year}{2021}).
	\newblock \urlprefix\url{https://doi.org/10.1103/PhysRevB.103.205127}.
	
	\bibitem{Lv2017}
	\bibinfo{author}{Lv, Y.-Y.} \emph{et~al.}
	\newblock \bibinfo{title}{{Experimental Observation of Anisotropic
			Adler-Bell-Jackiw Anomaly in Type-II Weyl Semimetal WTe$_{1.98}$ Crystals at
			the Quasiclassical Regime}}.
	\newblock \emph{\bibinfo{journal}{Phys. Rev. Lett.}}
	\textbf{\bibinfo{volume}{118}}, \bibinfo{pages}{096603}
	(\bibinfo{year}{2017}).
	\newblock \urlprefix\url{http://doi.org/10.1103/PhysRevLett.118.096603}.
	
	\bibitem{Zhang2016}
	\bibinfo{author}{Zhang, C.-L.} \emph{et~al.}
	\newblock \bibinfo{title}{{Signatures of the Adler–Bell–Jackiw chiral
			anomaly in a Weyl fermion semimetal}}.
	\newblock \emph{\bibinfo{journal}{Nat. Commun.}} \textbf{\bibinfo{volume}{7}},
	\bibinfo{pages}{10735} (\bibinfo{year}{2016}).
	\newblock \urlprefix\url{https://doi.org/10.1038/ncomms10735}.
	
	\bibitem{Ali2015}
	\bibinfo{author}{Ali, M.~N.} \emph{et~al.}
	\newblock \bibinfo{title}{{Correlation of crystal quality and extreme
			magnetoresistance of WTe 2}}.
	\newblock \emph{\bibinfo{journal}{EPL}} \textbf{\bibinfo{volume}{110}},
	\bibinfo{pages}{67002} (\bibinfo{year}{2015}).
	\newblock \urlprefix\url{https://doi.org/10.1209/0295-5075/110/67002}.
	
	\bibitem{zhu.singh.20}
	\bibinfo{author}{Zhu, Y.} \emph{et~al.}
	\newblock \bibinfo{title}{Exceptionally large anomalous hall effect due to
		anticrossing of spin-split bands in the antiferromagnetic half-heusler
		compound {TbPtBi}}.
	\newblock \emph{\bibinfo{journal}{Phys. Rev. B}}
	\textbf{\bibinfo{volume}{101}}, \bibinfo{pages}{161105}
	(\bibinfo{year}{2020}).
	\newblock \urlprefix\url{https://doi.org/10.1103/PhysRevB.101.161105}.
	
	\bibitem{Chen2021d}
	\bibinfo{author}{Chen, J.} \emph{et~al.}
	\newblock \bibinfo{title}{{Large anomalous Hall angle accompanying the sign
			change of anomalous Hall conductance in the topological half-Heusler compound
			HoPtBi}}.
	\newblock \emph{\bibinfo{journal}{Phys. Rev. B}}
	\textbf{\bibinfo{volume}{103}}, \bibinfo{pages}{144425}
	(\bibinfo{year}{2021}).
	\newblock \urlprefix\url{https://doi.org/10.1103/PhysRevB.103.144425}.
	
	\bibitem{souza.crivillero.23}
	\bibinfo{author}{Souza, J.~C.} \emph{et~al.}
	\newblock \bibinfo{title}{Tuning the topological character of half-heusler
		systems: A comparative study on {Y$T$Bi} ({$T=$Pd, Pt}}.
	\newblock \emph{\bibinfo{journal}{Phys. Rev. B}}
	\textbf{\bibinfo{volume}{108}}, \bibinfo{pages}{165154}
	(\bibinfo{year}{2023}).
	\newblock \urlprefix\url{https://doi.org/10.1103/PhysRevB.108.165154}.
	
	\bibitem{Feng2010}
	\bibinfo{author}{Feng, W.}, \bibinfo{author}{Xiao, D.}, \bibinfo{author}{Zhang,
		Y.} \& \bibinfo{author}{Yao, Y.}
	\newblock \bibinfo{title}{{Half-Heusler topological insulators: A
			first-principles study with the Tran-Blaha modified Becke-Johnson density
			functional}}.
	\newblock \emph{\bibinfo{journal}{Phys. Rev. B}} \textbf{\bibinfo{volume}{82}},
	\bibinfo{pages}{235121} (\bibinfo{year}{2010}).
	\newblock \urlprefix\url{http://doi.org/10.1103/PhysRevB.82.235121}.
	
	\bibitem{kozlova.hagel.05}
	\bibinfo{author}{Kozlova, N.} \emph{et~al.}
	\newblock \bibinfo{title}{Magnetic-field-induced band-structure change in
		{CeBiPt}}.
	\newblock \emph{\bibinfo{journal}{Phys. Rev. Lett.}}
	\textbf{\bibinfo{volume}{95}}, \bibinfo{pages}{086403}
	(\bibinfo{year}{2005}).
	\newblock \urlprefix\url{https://doi.org/10.1103/PhysRevLett.95.086403}.
	
	\bibitem{cano.bradlyn.17}
	\bibinfo{author}{Cano, J.} \emph{et~al.}
	\newblock \bibinfo{title}{Chiral anomaly factory: Creating {Weyl} fermions with
		a magnetic field}.
	\newblock \emph{\bibinfo{journal}{Phys. Rev. B}} \textbf{\bibinfo{volume}{95}},
	\bibinfo{pages}{161306} (\bibinfo{year}{2017}).
	\newblock \urlprefix\url{https://doi.org/10.1103/PhysRevB.95.161306}.
	
	\bibitem{burkov.14}
	\bibinfo{author}{Burkov, A.~A.}
	\newblock \bibinfo{title}{Anomalous {Hall} effect in {Weyl} metals}.
	\newblock \emph{\bibinfo{journal}{Phys. Rev. Lett.}}
	\textbf{\bibinfo{volume}{113}}, \bibinfo{pages}{187202}
	(\bibinfo{year}{2014}).
	\newblock \urlprefix\url{https://doi.org/10.1103/PhysRevLett.113.187202}.
	
	\bibitem{yao.kleinman.04}
	\bibinfo{author}{Yao, Y.} \emph{et~al.}
	\newblock \bibinfo{title}{First principles calculation of anomalous {Hall}
		conductivity in ferromagnetic bcc {Fe}}.
	\newblock \emph{\bibinfo{journal}{Phys. Rev. Lett.}}
	\textbf{\bibinfo{volume}{92}}, \bibinfo{pages}{037204}
	(\bibinfo{year}{2004}).
	\newblock \urlprefix\url{https://doi.org/10.1103/PhysRevLett.92.037204}.
	
	\bibitem{Pavlosiuk2016d}
	\bibinfo{author}{Pavlosiuk, O.}, \bibinfo{author}{Kaczorowski, D.} \&
	\bibinfo{author}{Wi{\'{s}}niewski, P.}
	\newblock \bibinfo{title}{{Magnetic and Transport Properties of Possibly
			Topologically Nontrivial Half-Heusler Bismuthides RMBi (R = Y, Gd, Dy, Ho,
			Lu; M = Pd, Pt)}}.
	\newblock \emph{\bibinfo{journal}{Acta Phys. Pol. A}}
	\textbf{\bibinfo{volume}{130}}, \bibinfo{pages}{573} (\bibinfo{year}{2016}).
	\newblock \urlprefix\url{https://doi.org/10.12693/APhysPolA.130.573}.
	
	\bibitem{Rodriguez-Carvajal1993}
	\bibinfo{author}{Rodr{\'{i}}guez-Carvajal, J.}
	\newblock \bibinfo{title}{{Recent advances in magnetic structure determination
			by neutron powder diffraction}}.
	\newblock \emph{\bibinfo{journal}{Physica B: Condensed Matter}}
	\textbf{\bibinfo{volume}{192}}, \bibinfo{pages}{55} (\bibinfo{year}{1993}).
	\newblock \urlprefix\url{https://doi.org/10.1016/0921-4526(93)90108-I}.
	
	\bibitem{Mott1929}
	\bibinfo{author}{Mott, N.~F.}
	\newblock \bibinfo{title}{{The scattering of fast electrons by atomic nuclei}}.
	\newblock \emph{\bibinfo{journal}{Proc. R. Soc. A}}
	\textbf{\bibinfo{volume}{124}}, \bibinfo{pages}{425} (\bibinfo{year}{1929}).
	\newblock \urlprefix\url{https://doi.org/10.1098/rspa.1929.0127}.
	
	\bibitem{Oen1973}
	\bibinfo{author}{Oen, O.~S.}
	\newblock \bibinfo{title}{{Cross sections for atomic displacements in solids by
			fast electrons, technical report, Office of Scientific and Technical
			Information (OSTI), ID No. 4457758, Report No. ORNL-4897}}
	(\bibinfo{year}{1973}).
	\newblock \urlprefix\url{https://www.osti.gov/biblio/4457758}.
	
	\bibitem{blochl.94}
	\bibinfo{author}{Bl\"ochl, P.~E.}
	\newblock \bibinfo{title}{Projector augmented-wave method}.
	\newblock \emph{\bibinfo{journal}{Phys. Rev. B}} \textbf{\bibinfo{volume}{50}},
	\bibinfo{pages}{17953} (\bibinfo{year}{1994}).
	\newblock \urlprefix\url{http://doi.org/10.1103/PhysRevB.50.17953}.
	
	\bibitem{kresse.hafner.94}
	\bibinfo{author}{Kresse, G.} \& \bibinfo{author}{Hafner, J.}
	\newblock \bibinfo{title}{Ab initio molecular-dynamics simulation of the
		liquid-metal--amorphous-semiconductor transition in germanium}.
	\newblock \emph{\bibinfo{journal}{Phys. Rev. B}} \textbf{\bibinfo{volume}{49}},
	\bibinfo{pages}{14251} (\bibinfo{year}{1994}).
	\newblock \urlprefix\url{http://doi.org/10.1103/PhysRevB.49.14251}.
	
	\bibitem{kresse.furthmuller.96}
	\bibinfo{author}{Kresse, G.} \& \bibinfo{author}{Furthm\"uller, J.}
	\newblock \bibinfo{title}{Efficient iterative schemes for ab initio
		total-energy calculations using a plane-wave basis set}.
	\newblock \emph{\bibinfo{journal}{Phys. Rev. B}} \textbf{\bibinfo{volume}{54}},
	\bibinfo{pages}{11169} (\bibinfo{year}{1996}).
	\newblock \urlprefix\url{http://doi.org/10.1103/PhysRevB.54.11169}.
	
	\bibitem{kresse.joubert.99}
	\bibinfo{author}{Kresse, G.} \& \bibinfo{author}{Joubert, D.}
	\newblock \bibinfo{title}{From ultrasoft pseudopotentials to the projector
		augmented-wave method}.
	\newblock \emph{\bibinfo{journal}{Phys. Rev. B}} \textbf{\bibinfo{volume}{59}},
	\bibinfo{pages}{1758} (\bibinfo{year}{1999}).
	\newblock \urlprefix\url{http://doi.org/10.1103/PhysRevB.59.1758}.
	
	\bibitem{becke.johnson.06}
	\bibinfo{author}{Becke, A.~D.} \& \bibinfo{author}{Johnson, E.~R.}
	\newblock \bibinfo{title}{A simple effective potential for exchange}.
	\newblock \emph{\bibinfo{journal}{J. Chem. Phys.}}
	\textbf{\bibinfo{volume}{124}}, \bibinfo{pages}{221101}
	(\bibinfo{year}{2006}).
	\newblock \urlprefix\url{https://doi.org/10.1063/1.2213970}.
	
	\bibitem{tran.blaha.09}
	\bibinfo{author}{Tran, F.} \& \bibinfo{author}{Blaha, P.}
	\newblock \bibinfo{title}{Accurate band gaps of semiconductors and insulators
		with a semilocal exchange-correlation potential}.
	\newblock \emph{\bibinfo{journal}{Phys. Rev. Lett.}}
	\textbf{\bibinfo{volume}{102}}, \bibinfo{pages}{226401}
	(\bibinfo{year}{2009}).
	\newblock \urlprefix\url{https://doi.org/10.1103/PhysRevLett.102.226401}.
	
	\bibitem{camargo.baquero.12}
	\bibinfo{author}{Camargo-Mart\'{\i}nez, J.~A.} \& \bibinfo{author}{Baquero, R.}
	\newblock \bibinfo{title}{Performance of the modified {Becke-Johnson} potential
		for semiconductors}.
	\newblock \emph{\bibinfo{journal}{Phys. Rev. B}} \textbf{\bibinfo{volume}{86}},
	\bibinfo{pages}{195106} (\bibinfo{year}{2012}).
	\newblock \urlprefix\url{https://doi.org/10.1103/PhysRevB.86.195106}.
	
	\bibitem{monkhorst.pack.76}
	\bibinfo{author}{Monkhorst, H.~J.} \& \bibinfo{author}{Pack, J.~D.}
	\newblock \bibinfo{title}{Special points for {Brillouin}-zone integrations}.
	\newblock \emph{\bibinfo{journal}{Phys. Rev. B}} \textbf{\bibinfo{volume}{13}},
	\bibinfo{pages}{5188} (\bibinfo{year}{1976}).
	\newblock \urlprefix\url{http://doi.org/10.1103/PhysRevB.13.5188}.
	
	\bibitem{pizzi.vitale.20}
	\bibinfo{author}{Pizzi, G.} \emph{et~al.}
	\newblock \bibinfo{title}{Wannier90 as a community code: new features and
		applications}.
	\newblock \emph{\bibinfo{journal}{J. Phys.: Condens. Matter}}
	\textbf{\bibinfo{volume}{32}}, \bibinfo{pages}{165902}
	(\bibinfo{year}{2020}).
	\newblock \urlprefix\url{https://doi.org/10.1088/1361-648X/ab51ff}.
	
\end{thebibliography}

\begin{thebibliography}{1}
	\expandafter\ifx\csname url\endcsname\relax
	\def\url#1{\texttt{#1}}\fi
	\expandafter\ifx\csname urlprefix\endcsname\relax\def\urlprefix{URL }\fi
	\providecommand{\bibinfo}[2]{#2}
	\providecommand{\eprint}[2][]{\url{#2}}
	
	\bibitem{Zhu2023b_1}
	\bibinfo{author}{Zhu, Y.} \emph{et~al.}
	\newblock \bibinfo{title}{{Large anomalous Hall effect and negative
			magnetoresistance in half-topological semimetals}}.
	\newblock \emph{\bibinfo{journal}{Commun. Phys.}} \textbf{\bibinfo{volume}{6}},
	\bibinfo{pages}{346} (\bibinfo{year}{2023}).
	\newblock \urlprefix\url{https://doi.org/10.1038/s42005-023-01469-6}.
	
	\bibitem{Shekhar2018_1}
	\bibinfo{author}{Shekhar, C.} \emph{et~al.}
	\newblock \bibinfo{title}{{Anomalous Hall effect in Weyl semimetal half-Heusler
			compounds RPtBi (R = Gd and Nd)}}.
	\newblock \emph{\bibinfo{journal}{Proc. Natl. Acad. Sci. U.S.A.}}
	\textbf{\bibinfo{volume}{115}}, \bibinfo{pages}{9140} (\bibinfo{year}{2018}).
	\newblock \urlprefix\url{https://doi.org/10.1073/pnas.1810842115}.
	
	\bibitem{Suzuki2016_1}
	\bibinfo{author}{Suzuki, T.} \emph{et~al.}
	\newblock \bibinfo{title}{{Large anomalous Hall effect in a half-Heusler
			antiferromagnet}}.
	\newblock \emph{\bibinfo{journal}{Nat. Phys.}} \textbf{\bibinfo{volume}{12}},
	\bibinfo{pages}{1119} (\bibinfo{year}{2016}).
	\newblock \urlprefix\url{https://doi.org/10.1038/nphys3831}.
	
	\bibitem{Pavlosiuk2020_1}
	\bibinfo{author}{Pavlosiuk, O.}, \bibinfo{author}{Fa{\l}at, P.},
	\bibinfo{author}{Kaczorowski, D.} \& \bibinfo{author}{Wi{\'{s}}niewski, P.}
	\newblock \bibinfo{title}{{Anomalous Hall effect and negative longitudinal
			magnetoresistance in half-Heusler topological semimetal candidates TbPtBi and
			HoPtBi}}.
	\newblock \emph{\bibinfo{journal}{APL Mater.}} \textbf{\bibinfo{volume}{8}},
	\bibinfo{pages}{111107} (\bibinfo{year}{2020}).
	\newblock \urlprefix\url{https://doi.org/10.1063/5.0026956}.
	
	\bibitem{zhu.singh.20_1}
	\bibinfo{author}{Zhu, Y.} \emph{et~al.}
	\newblock \bibinfo{title}{Exceptionally large anomalous hall effect due to
		anticrossing of spin-split bands in the antiferromagnetic half-heusler
		compound {TbPtBi}}.
	\newblock \emph{\bibinfo{journal}{Phys. Rev. B}}
	\textbf{\bibinfo{volume}{101}}, \bibinfo{pages}{161105}
	(\bibinfo{year}{2020}).
	\newblock \urlprefix\url{https://doi.org/10.1103/PhysRevB.101.161105}.
	
	\bibitem{Chen2021d1}
	\bibinfo{author}{Chen, J.} \emph{et~al.}
	\newblock \bibinfo{title}{{Large anomalous Hall angle accompanying the sign
			change of anomalous Hall conductance in the topological half-Heusler compound
			HoPtBi}}.
	\newblock \emph{\bibinfo{journal}{Phys. Rev. B}}
	\textbf{\bibinfo{volume}{103}}, \bibinfo{pages}{144425}
	(\bibinfo{year}{2021}).
	\newblock \urlprefix\url{https://doi.org/10.1103/physrevb.103.144425}.
	
	\bibitem{Singha2019a}
	\bibinfo{author}{Singha, R.}, \bibinfo{author}{Roy, S.},
	\bibinfo{author}{Pariari, A.}, \bibinfo{author}{Satpati, B.} \&
	\bibinfo{author}{Mandal, P.}
	\newblock \bibinfo{title}{{Magnetotransport properties and giant anomalous Hall
			angle in the half-Heusler compound TbPtBi}}.
	\newblock \emph{\bibinfo{journal}{Phys. Rev. B}} \textbf{\bibinfo{volume}{99}},
	\bibinfo{pages}{035110} (\bibinfo{year}{2019}).
	\newblock \urlprefix\url{https://doi.org/10.1103/PhysRevB.99.035110}.
	
\end{thebibliography}

\end{document}